\documentclass[twocolumn, trackchanges]{aastex62}

\pdfoutput=1

\usepackage{amsmath,amssymb,amstext}
\usepackage{hyperref}  
\usepackage{graphicx} 
\usepackage{bm}
\usepackage{multirow}
\usepackage{gensymb}
\turnoffedit

\graphicspath{ {figures/} }

\shorttitle{GMC Collisions as Triggers of Star Formation. VII.}
\shortauthors{Wu et al.}

\begin{document}

\title{GMC Collisions as Triggers of Star Formation. VII. \\
The Effect of Magnetic Field Strength on Star Formation
}

\author{Benjamin Wu}
\affil{National Astronomical Observatory of Japan, Mitaka, Tokyo 181-8588, Japan}
\email{ben.wu@nao.ac.jp}
\author{Jonathan C. Tan}
\affil{Department of Space, Earth \& Environment, Chalmers University of Technology, SE-412 96 Gothenburg, Sweden}
\affil{Department of Astronomy, University of Virginia, Charlottesville, VA 22904, USA}
\author{Duncan Christie}
\affil{Department of Astronomy, University of Virginia, Charlottesville, VA 22904, USA}
\author{Fumitaka Nakamura}
\affil{National Astronomical Observatory of Japan, Mitaka, Tokyo 181-8588, Japan}

\begin{abstract}
We investigate the formation of stars within giant molecular clouds
(GMCs) evolving in environments of different global magnetic field
strength and large-scale dynamics. Building upon a series of
magnetohydrodynamic (MHD) simulations of non-colliding and colliding
GMCs, we employ density- and magnetically-regulated star formation
sub-grid models in clouds which range from moderately magnetically 
supercritical to near critical.
We examine gas and
star cluster morphologies, magnetic field strengths and relative
orientations, pre-stellar core densities, temperatures, mass-to-flux
ratios and velocities, star formation rates and efficiencies over
time, spatial clustering of stars, and kinematics of the stars and
natal gas. The large scale magnetic criticality of the region greatly
affects the overall gas evolution and star formation properties. GMC
collisions enhance star formation rates and efficiencies in
magnetically supercritical conditions, but may actually inhibit them
in the magnetically critical case.  This may have implications for
star formation in different Galactic environments such as the Galactic
Center and the main Galactic disk.
\end{abstract}


\section{Introduction}
\label{sec:intro}

Giant molecular cloud (GMC) collisions have been posited as a
mechanism for triggering the formation of stars and possibly even
setting global star formation rates (SFRs) in disk galaxies
\citep[e.g.,][]{Scoville_Sanders_Clemens_1986,Tan_2000}. Such
converging molecular flows are likely to form regions of dense,
gravitationally unstable gas with properties similar to those observed
in Infrared Dark Clouds \citep[IRDCs; see e.g.,][]{Tan_ea_2014}. 
\edit1{Such dense clumps are potential precursors to massive stars
and star clusters \citep[see
  e.g.,][]{Sanhueza_ea_2019,Moser_ea_2019}.}

On the other hand, magnetic fields ($B$-fields) in the interstellar
medium (ISM) may act as an important regulator of star formation in
GMCs, as studies have found reductions of overall fragmentation, SFRs,
and star formation efficiency (SFE) by factors of a few upon the
inclusion of $B$-fields in driven turbulence simulations \citep[see,
  e.g., ][]{Padoan_Nordlund_2011,Federrath_Klessen_2012}. $B$-fields in
conjunction with turbulence may help explain the very inefficient SFRs
per local free-fall time observed on average within GMCs
\citep{Zuckerman_Evans_1974, Krumholz_Tan_2007}, as thermal pressures
are relatively insignificant at the $T\sim 10-20\:$K temperatures within
GMCs. Such regulating mechanisms alone, however, may be unable to
explain the high variation in observed SFRs \citep{Lee_ea_2016}. Here,
irregular, intermittent phenomena such as GMC collisions may play a
key role.

\edit1{Previous numerical studies have shown that magnetized converging 
\textit{atomic} flows can create dense molecular clouds if the flows are 
initially magnetically supercritical, (i.e., the dimensionless mass-to-flux
ratio $\lambda> 1$). Converging atomic flows that 
are initially subcritical ($\lambda<1$), however, do not readily form 
supercritical structures even when including ambipolar diffusion 
\citep{Vazquez-Semadeni_ea_2011,Koertgen_Banerjee_2015}.
In cases of magnetized converging \textit{molecular} flows which are
already supercritical, the formation and subsequent collapse of 
supercritical cores occurs rapidly within the shock-compressed dense layers \citep{Chen_Ostriker_2014}.
}

\edit1{However,} 
the frequency of cloud-cloud interactions has been difficult to
determine. Earlier studies estimated timescales of order 100~Myr
between collisions, casting doubt on the prevalence of this mechanism
\citep{Blitz_Shu_1980}. However, upon accounting for self-gravity,
differential rotation, and an effectively 2D geometry due to disk
scale height constraints, predicted cloud collision timescales are
reduced significantly \citep{Gammie_ea_1991,Tan_2000}.  Global galaxy
simulations by, e.g.,
\citet{Tasker_Tan_2009,Dobbs_ea_2015,Fujimoto_ea_2014,Li_ea_2018}
showed that, on average, a molecular cloud experiences a collision
every $\sim1/5$ of a local 
galactic orbit (i.e., every $\sim$20~Myr at a galactocentric radius of
$\sim4\:$kpc in the Milky Way), with even shorter timescales possible
for the most massive clouds and in the presence of spiral and bar
potentials.
\edit1{The inclusion of $B$-fields in global galaxy simulations has been
shown to produce regions with wide-ranging levels of magnetic criticality, 
some of which are favorable for the formation and accumulation of GMCs \citep{Koertgen_ea_2018,Koertgen_ea_2019}.}

A growing number of observations of dense clumps and young massive
clusters have claimed evidence of cloud-cloud collisions \citep[e.g.,
][]{Loren_1976,Furukawa_ea_2009,Torii_ea_2011,Nakamura_ea_2012,Sanhueza_ea_2013,Fukui_ea_2014,Nishimura_ea_2018,Dobashi_ea_2019}. 
\edit1{These are often compared against signatures predicted from 
numerical simulations of colliding molecular clouds \citep[e.g., ][]{Habe_Ohta_1992,Klein_Woods_1998,Anathpindika_2009,Takahira_ea_2014,Haworth_ea_2015a,Balfour_ea_2015,Wu_ea_2017a,Wu_ea_2017b,Bisbas_ea_2017,Wu_ea_2018}.} 
Disrupted
gas morphology, multiple velocity components, proximity to young
massive stars, broad bridge features in position-velocity space, 
and molecular and atomic kinematic tracers are all diagnostics predicted to
differentiate GMC collisions from non-colliding clouds. Frequently,
however, the complexity of the blue and redshifted velocity fields and
the ambiguity of observational features cannot rule out alternative
explanations, such as internal gas motions or coincidental projection
effects.

Nevertheless, the nature of typical cloud collisions, their outcomes,
and definitive ways to distinguish them are all still unanswered
questions.  The current work continues a series of papers that has
been methodically studying the nature of GMC-GMC collisions within a
magnetized ISM. Papers I \citep{Wu_ea_2015} and II \citep{Wu_ea_2017a}
performed parameter space explorations and laid the numerical
framework in 2D and 3D, respectively, of ideal magnetohydrodynamics
(MHD), gas heating/cooling, and turbulence. Paper III
\citep{Wu_ea_2017b} implemented star formation in the form of two
sub-grid models: density regulated and magnetically regulated. Paper
IV \citep{Christie_ea_2017} implemented the non-ideal MHD effects of
ambipolar diffusion, Paper V \citep{Bisbas_ea_2017} examined
observational signatures using radiative transfer post-processing, and
Paper VI \citep{Wu_ea_2018} studied collision-induced turbulence.

In this paper, we investigate how the strength of the magnetic field
affects the nature of star formation in non-colliding and colliding
GMCs. We aim to expand our understanding of how star formation
proceeds in different galactic environments, where
$10^{5}\:M_{\odot}$, $r\sim1$pc clouds can be observed to be mostly
starless (e.g., in the Galactic Center) or forming massive young
clusters (e.g., in the disk) \citep[see,
  e.g.,][]{Longmore_ea_2014,Tan_ea_2014,Sanhueza_ea_2019}.

Our numerical set-up is presented in
\S\ref{sec:methods}. \S\ref{sec:results} details the
results, which include cloud and cluster morphologies, magnetic field
orientations and strengths, probability distribution functions (PDFs),
properties of star-forming gas, star formation rates and efficiencies, 
spatial clustering, and star versus gas kinematics. 
Finally, conclusions are discussed in \S\ref{sec:conclusion}.

\section{Numerical Model}
\label{sec:methods}

\subsection{Initial Conditions}

Our numerical simulations are based on the GMC models introduced in
Paper II with star-formation routines introduced in Paper III and a
number of numerical improvements from Paper IV. Specifically, we
include heating/cooling, self-gravity, supersonic turbulence, ideal
MHD, and investigate both density and magnetically-regulated star
formation. In this work, our main goal is to explore the effects of
varying magnetic field strengths at the global GMC scale. This, in
turn, affects the star formation routines at the sub-grid scale. The
physical properties of our models are summarized in
Table~\ref{tab:initial} and detailed below.

Within a simulation cube of side length $L=128\:{\rm pc}$, two
identical GMCs are initialized as uniform spheres with radii $R_{\rm
  GMC} = 20.0$~pc and Hydrogen number densities $n_{\rm H,GMC} =
100\:{\rm cm^{-3}}$, giving masses of $M_{\rm GMC} = 9.3 \times
10^4\:M_\odot$. The GMC centers are offset by an impact parameter $b =
0.5 R_{\rm GMC}$. The clouds are embedded within ambient gas of ten
times lower density $n_{\rm H,0} = 10\:{\rm cm^{-3}}$, filling the
remainder of the volume and representative of an atomic cold neutral
medium (CNM). In the non-colliding model, there is no additional bulk
velocity field, while in the colliding model, both the CNM and GMCs
are converging with a relative velocity of $v_{\rm rel}=10\:{\rm
  km\:s^{-1}}$ along the collision axis (defined as the
$x$-axis). This collision velocity is consistent with the peak of the
distribution of relative velocities from interacting GMCs tracked in
global galactic simulations \citep[e.g.,][]{Li_ea_2018}.

A uniform magnetic field is initialized throughout the entire domain
at an angle $\theta=60^{\circ}$ with respect to the collision axis. To
investigate the role of magnetic strength on gas dynamics and star
formation, we initialize fields with magnitudes of $B=10, 30,$ and
$50\:{\rm \mu G}$, which correspond to configurations of GMCs with
average dimensionless mass-to-flux ratios $\lambda_{\rm
  GMC}=(M/\Phi)(\sqrt{G}/0.126)$ = 5.4 (i.e., moderately
magnetically supercritical), 1.8 (i.e., marginally supercritical), and
1.1 (i.e., near critical), respectively. 
Note that due to the spherical geometry, there are much larger columns through the cloud centers, resulting in higher degrees of supercriticality in central flux tubes and lower degrees near cloud boundaries.
The equilibrium
temperature of molecular gas within the GMCs is $\sim 15$~K, which
yields thermal-to-magnetic pressure ratios $\beta=8\pi
c_{s}^{2}\rho/B^{2}=1.5\times10^{-2}$, $1.6\times10^{-3}$, and
$6.0\times10^{-4}$, respectively. While $B\sim 10\:{\rm \mu G}$ has
been inferred from Zeeman measurements of nearby GMCs, as summarized
by \citet{Crutcher_2012}, much stronger magnetic fields of order
\edit1{$\sim0.1-1\:{\rm mG}$} have been estimated to be present in IRDCs
\citep[e.g.,][]{Pillai_ea_2015,Pillai_ea_2016,LiuT_ea_2018,Soam_ea_2019}.

We approximate the complex density and velocity structures observed in
GMCs by initializing gas within our model clouds with a
solenoidal random supersonic turbulent velocity field. The
clouds are of order virial, with Mach number $\mathcal{M}_{s} \equiv
\sigma/ c_{s}= 23$ (for $T=15$~K). The velocity field follows a
$v_{k}^{2} \propto k^{-4}$ relation, where $k$ is the wavenumber and
each $k$-mode spanning $2<k/(\pi/L)<20$ is excited. We do not drive
turbulence, but rather let it decay. 
Note, however, that GMC collisions themselves provide an
additional mode of turbulence driving \citep{Wu_ea_2018}.

As in Paper III, the simulations are run for $5\:{\rm Myr}$ to
investigate the initial phases of star formation in both non-colliding
and colliding cases. For reference, the freefall time for the initial
uniform density GMCs is $t_{\rm ff} = (3\pi/[32G\rho])^{1/2} = 4.35$
Myr, but the values of the local $t_{\rm ff}$ for denser substructures
formed from turbulence and the collision are much shorter.

\begin{deluxetable}{rlcc}
\tablecaption{Initial Simulation Properties \label{tab:initial}}
\tablecolumns{4}
\tablewidth{0pt}
\tablehead{
\multicolumn{2}{l}{Gas Properties} & \colhead{GMC} &\colhead{Ambient}
}
\startdata
$n_{\rm H}$ & (${\rm cm}^{-3}$)  & 100                   & 10      \\
$R$         & (${\rm pc}$)       & 20                    &  -      \\
$M$         & ($M_{\odot}$)      & $9.3\times10^{4}$     &  -      \\
$t_{\rm ff}$ & (Myr)             & 4.35                  & 13.8    \\
$T$         & (K)                & 15                    & 150     \\
$c_{\rm s}$ & (km/s)             & 0.23                  & 0.72    \\
$v_{\rm vir}$ & (km/s)           & 4.9                   &  -      \\
$v_{\rm bulk}$ & (km/s)          & \multicolumn{2}{c}{[0; $\pm5$]\tablenotemark{a}}  \\
\hline
\multicolumn{4}{l}{Turbulence Properties} \\
\hline
$k$-mode    &  ($\pi/L$)                  & $\{ k_{1}=2,k_{2}=20 \}$               &  -      \\
$\sigma$    & (km/s)             & 5.2                   &  -      \\
$\mathcal{M}_{\rm s}$ &          & 23                    & -       \\
\hline
\multicolumn{4}{l}{Magnetic Field Properties}  \\
\hline
$B$         & (${\rm \mu G}$)    & \multicolumn{2}{c}{(10, 30, 50)\tablenotemark{a}}  \\
$\beta$\tablenotemark{b}     &                    & \multicolumn{2}{c}{($1.5\times10^{-2}$, $1.7\times10^{-3}$, $6.0\times10^{-4}$)\tablenotemark{a}} \\
$\lambda$\tablenotemark{c}   &                    & (5.8, 1.8, 1.1)      & (1.9, 0.6, 0.4)       \\
$v_{A}$     & (km/s)             & (1.84, 5.52, 9.2)     & (5.83, 17.49, 29.2)    \\
$\mathcal{M}_{A}$ &              & (2.82, 0.94, 0.57)    & -       \\ 
\enddata

\tablenotetext{a}{applies to both regions}
\tablenotetext{b}{thermal-to-magnetic pressure ratio: $\beta=8\pi c_{s}^2 \rho_{0} / B^2$}
\tablenotetext{c}{normalized mass-to-flux ratio: $\lambda = (M/\Phi)(\sqrt{G}/0.126)$}
\end{deluxetable}

\subsection{Numerical Code}
\label{sec:methods-code}

The magnetohydrodynamics (MHD) adaptive mesh refinement (AMR) code
\texttt{Enzo}\footnote{http://enzo-project.org (v2.4)}
\citep{Bryan_ea_2014} is used to run our simulations, specifically
utilizing the Dedner MHD solver and hyperbolic divergence cleaning
method \citep{Dedner_ea_2002,Wang_Abel_2008}. We use a root grid of
$128^{3}$ with 3 additional levels of AMR, resulting in an effective
resolution of $1024^{3}$ and minimum grid cell size of 0.125~pc.
While our refinement criterion is based on resolving the Jeans length
by 8 cells \citep[see, e.g.,][]{Truelove_ea_1997}, this means that 
finer grid cells are placed everywhere in the GMC
regions and Jeans fragmentation is not well resolved at densities
where the Jeans length becomes $\lesssim 1\:$pc 
(i.e., $n_{\rm H} \gtrsim 2.1\times 10^{2}\:{\rm cm^{-3}}$). However, note that
the Jeans criterion assumes purely thermal support, so the effective
``magneto-Jeans length'' of magnetized gas will be significantly
larger perpendicular to the magnetic field, so magnetically-regulated
fragmentation will be better resolved. Still, it is not our goal in
this paper to accurately follow the fragmentation of very dense gas
structures, e.g., that may be relevant to the core mass function, but
rather the overall efficiency of dense gas formation and more global
aspects of its morphology.

Heating and cooling are governed by functions based on modeling of
photo-dissociation regions \citep[PDRs; see][]{Wu_ea_2015} and
implemented via the Grackle 2.2 chemistry library
\citep{Smith_ea_2017}. We additionally utilize: the ``dual energy
formalism'' \citep{Bryan_ea_2014}, important for accurate calculations
of pressures and temperatures in conditions of low thermal-to-total
energy ratios (e.g., in the presence of relatively high bulk
velocities and strong magnetic fields); an ``Alfv\'en limiter''
(described in Paper II) to avoid extremely short timesteps set by
Alfv\'en waves by choosing a maximum Alfv\'en velocity, $v_{A,{\rm
    max}}=B/\sqrt{4\pi\rho_{\rm min}}=1\times10^{7}\:{\rm
  cm\:s^{-1}}$; and a minimum cooling timestep (described in
Paper IV) of $t_{\rm cool,min}\sim 600$~yr to avoid prohibitively
short timesteps at high densities.

\subsection{Star Formation Models}
\label{sec:methods-SF}

The star formation process is represented by the density-regulated and
magnetically-regulated sub-grid star formation routines introduced in
Paper III. For both star formation models, star particles (i.e.,
collisionless, point particles with mass $m_{\star}$) form within a
simulation cell if the following criteria are met. First, the cell
must be at the finest level of resolution, determined by the local
Jeans length. Second, the temperature in the cell must be $<3000$~K to
prevent stars from forming in transient shock-heated dense
regions. Third, the cell must exceed certain physical thresholds.  In
the density-regulated model, a constant density threshold of $n_{\rm
  H,sf}=1.0\times 10^{6}\:{\rm cm}^{-3}$ is required.  In the
magnetically-regulated model, the cell must be locally magnetically
``supercritical'' (i.e., having a mass-to-flux ratio $\mu_{\rm
  cell}>1$). The dimensionless mass-to-flux ratio is
\begin{equation}
\mu_{\rm cell} = \frac{\rho \Delta x \sqrt{G}}{B c_{1}}
\end{equation}
for a cell of edge length $\Delta x$, gas density $\rho$, magnetic
field strength $B$, and using gravitational constant $G$.  The value
of $c_{1}$ depends on the extended geometry of the flux tube, which we
set to be $c_{\rm 1,fid}=0.126$ \citep[for an isolated cloud,
][]{Mouschovias_Spitzer_1976}. See also \citet{Nakano_Nakamura_1978}
for an infinite disk, in which case
$c_{1}=\frac{1}{2\pi}\sim0.159$. We explore variations in this
threshold of a factor of two higher and lower for the
magnetically-regulated model.

The resulting expression for star-formation threshold density as a
function of magnetic field strength can then be written as
\begin{eqnarray}
n_{\rm H} & > & \frac{B c_{1}}{\mu_{\rm H} m_{\rm H} \Delta x \sqrt{G}} \\ & = & 5.42 \times 10^{3} \left(\frac{B}{10\:{\rm \mu G}}\right) \left(\frac{c_{1}}{0.126}\right)\left(\frac{\Delta x}{0.125\:{\rm pc}}\right)^{-1} {\rm cm}^{-3},
\end{eqnarray}
where the mean particle mass per hydrogen is $\mu_{\rm H} m_{\rm H}
=1.4 m_{\rm H}$.

Finally, if the required thresholds for a given model are satisfied, a
fraction of the gas mass in the cell is transformed into star
particles such that the average rate follows a fixed star formation
efficiency per local free-fall time, $\epsilon_{\rm ff}=0.02$.  The
local free-fall time, $t_{\rm ff}$, approximated as the collapse of a
uniform density sphere, is
\begin{eqnarray}
t_{\rm ff} & = & \left(\frac{3\pi}{32G\rho}\right)^{1/2} \\ & = & 4.4\times10^{4} n_{\rm H,6}^{-1/2}\:{\rm yr},
\end{eqnarray}
where
$n_{\rm H,6}\equiv n_{\rm H}/10^6\:{\rm cm^{-3}}$. 
This results in a SFR of
\begin{eqnarray}\label{eq:SFRmdot}
\dot{m}_\star & = & \epsilon_{\rm ff}\frac{m_{\rm gas}}{t_{\rm ff}} \\ & = & 2.9\times10^{-5} \left(\frac{\epsilon_{\rm ff}}{0.02}\right)\left(\frac{\Delta x}{0.125\:{\rm pc}}\right)^3 n_{\rm H,6}^{3/2}\:M_\odot \:{\rm yr}^{-1}.
\end{eqnarray}

\begin{table*}
  \caption{List of Simulations}
  \label{tab:list_of_sims}
  \centering
  \begin{tabular}{lccccccccc} \hline \hline
Name & Star Formation & $v_{\rm rel}$       & $B$        & $\Delta x$ & $n_{\rm H,sf}$ & $t_{\rm ff}$ & $m_{\rm gas,min}$ & $m_{\rm \star, min}$ &  $c_{1}$  \\
     & Model          &(${\rm km\:s^{-1}}$) & ($\mu$G) & (pc)       &  (cm$^{-3}$)   & (years)      &  ($M_{\odot}$)    &   ($M_{\odot}$)      & ($c_{\rm 1,fid}$)      \\
\hline
B10-d1-nocol & density  & 0 & 10 & 0.125 &  $1.0\times10^{6}$ & $4.4\times 10^{4}$ & 6.3 & 1 & ...   \\
B10-05-nocol & magnetic & 0 & 10 & 0.125 & $3.55\times10^{4}$ & $2.3\times 10^{5}$ &  2  & 1 & $0.5$ \\
B10-1-nocol  & magnetic & 0 & 10 & 0.125 & $3.55\times10^{4}$ & $2.3\times 10^{5}$ &  2  & 1 & $1.0$ \\
B10-2-nocol  & magnetic & 0 & 10 & 0.125 & $3.55\times10^{4}$ & $2.3\times 10^{5}$ &  2  & 1 & $2.0$ \\
B30-d1-nocol & density  & 0 & 30 & 0.125 &  $1.0\times10^{6}$ & $4.4\times 10^{4}$ & 6.3 & 1 &  ...  \\
B30-05-nocol & magnetic & 0 & 30 & 0.125 & $3.55\times10^{4}$ & $2.3\times 10^{5}$ &  2  & 1 & $0.5$ \\
B30-1-nocol  & magnetic & 0 & 30 & 0.125 & $3.55\times10^{4}$ & $2.3\times 10^{5}$ &  2  & 1 & $1.0$ \\
B30-2-nocol  & magnetic & 0 & 30 & 0.125 & $3.55\times10^{4}$ & $2.3\times 10^{5}$ &  2  & 1 & $2.0$ \\
B50-d1-nocol & density  & 0 & 50 & 0.125 &  $1.0\times10^{6}$ & $4.4\times 10^{4}$ & 6.3 & 1 &  ...  \\
B50-05-nocol & magnetic & 0 & 50 & 0.125 & $3.55\times10^{4}$ & $2.3\times 10^{5}$ &  2  & 1 & $0.5$ \\
B50-1-nocol  & magnetic & 0 & 50 & 0.125 & $3.55\times10^{4}$ & $2.3\times 10^{5}$ &  2  & 1 & $1.0$ \\
B50-2-nocol  & magnetic & 0 & 50 & 0.125 & $3.55\times10^{4}$ & $2.3\times 10^{5}$ &  2  & 1 & $2.0$ \\
\hline
B10-d1-col   & density  &10 & 10 & 0.125 &  $1.0\times10^{6}$ & $4.4\times 10^{4}$ & 6.3 & 1 &  ...  \\
B10-05-col   & magnetic &10 & 10 & 0.125 & $3.55\times10^{4}$ & $2.3\times 10^{5}$ &  2  & 1 & $0.5$ \\
B10-1-col    & magnetic &10 & 10 & 0.125 & $3.55\times10^{4}$ & $2.3\times 10^{5}$ &  2  & 1 & $1.0$ \\
B10-2-col    & magnetic &10 & 10 & 0.125 & $3.55\times10^{4}$ & $2.3\times 10^{5}$ &  2  & 1 & $2.0$ \\
B30-d1-col   & density  &10 & 30 & 0.125 &  $1.0\times10^{6}$ & $4.4\times 10^{4}$ & 6.3 & 1 &  ...  \\
B30-05-col   & magnetic &10 & 30 & 0.125 & $3.55\times10^{4}$ & $2.3\times 10^{5}$ &  2  & 1 & $0.5$ \\
B30-1-col    & magnetic &10 & 30 & 0.125 & $3.55\times10^{4}$ & $2.3\times 10^{5}$ &  2  & 1 & $1.0$ \\
B30-2-col    & magnetic &10 & 30 & 0.125 & $3.55\times10^{4}$ & $2.3\times 10^{5}$ &  2  & 1 & $2.0$ \\
B50-d1-col   & density  &10 & 50 & 0.125 &  $1.0\times10^{6}$ & $4.4\times 10^{4}$ & 6.3 & 1 &  ...  \\
B50-05-col   & magnetic &10 & 50 & 0.125 & $3.55\times10^{4}$ & $2.3\times 10^{5}$ &  2  & 1 & $0.5$ \\
B50-1-col    & magnetic &10 & 50 & 0.125 & $3.55\times10^{4}$ & $2.3\times 10^{5}$ &  2  & 1 & $1.0$ \\
B50-2-col    & magnetic &10 & 50 & 0.125 & $3.55\times10^{4}$ & $2.3\times 10^{5}$ &  2  & 1 & $2.0$ \\
\hline
\end{tabular}
\end{table*}

As our simulation timesteps are much shorter than the sound crossing
time for a cell (i.e., $\ll 1.2\times 10^{5}$ yr for a $1\:{\rm
  km\:s^{-1}}$ signal speed), the rates defined above typically lead
to small expected stellar masses ($\lesssim 1\:M_\odot$). However, to
avoid excessively large numbers of star particles we adopt a minimum
star particle mass, $m_{\rm \star,min} = 1\:M_\odot$. If the stellar
mass expected to be formed in a cell is $m_{\rm \star} < 1\:M_\odot$,
its formation is treated stochastically. In this ``stochastic
regime'', a $1\:M_\odot$ star particle is created with a probability
$\dot{m}_\star \Delta t / m_{\rm \star,min}$, where $\Delta t$ is the
simulation timestep. However, for cases where $\dot{m}_\star \Delta t
> m_{\rm \star,min}$, a star particle with this mass is simply formed
and it is possible for a distribution of initial stellar masses to be
created. However, we caution that with such a simple model we do not
expect this distribution to necessarily have any similarity to that of
the actual stellar initial mass function (IMF).

Note that we limit the fraction of gas mass in a cell that can turn
into stars within a single timestep to $<0.5$.
In conjunction with the minimum star particle mass, this imposes an
effective density threshold on the cell of $n_{\rm H,sf}= 3.55\times
10^{4}\:{\rm cm^{-3}}$. This threshold plays a role in the
magnetically-regulated SF routine but is superseded by the standard
density threshold of the density-regulated SF routine.

\section{Results}
\label{sec:results}

Overall, we compare the results of 24 different simulations. For both
non-colliding (``nocol'') and colliding (``col'') GMC cases, magnetic
field strengths of 10, 30, and 50$\mu$G are initialized,
respectively. For each of these cases, one density regulated and three
magnetically regulated SF models ($c_{1}$=0.5, 1.0, and 2.0 $c_{\rm 1,
  fid}$) are run. The star formation and physical parameters used each
simulation are listed in Table~\ref{tab:list_of_sims}.

We analyze various aspects of star formation activity among the
collection of non-colliding and colliding simulations at different
magnetic field strengths.  In particular, we discuss: morphology of
the clouds and clusters (\S\ref{sec:results-morph}); magnetic field
orientations and strengths (\S\ref{sec:results-Bfields}); probability
distribution functions (\S\ref{sec:results-PDFs}); properties
of star-forming gas (\S\ref{sec:results-SFcells}); global star
formation rates (\S\ref{sec:results-SFR}); spatial clustering of stars
(\S\ref{sec:results-clustering}); and star versus gas kinematics
(\S\ref{sec:results-PV}).

In most visualizations, we use the $(x,y,z)$ coordinate system as
defined in the simulations. Occasionally, we use $(x',y',z')$, where 
the axes are rotated by the polar and azimuthal angles 
$(\theta, \phi)=(15^{\circ},15^{\circ})$ when appropriate 
(e.g., representing certain observables).

\subsection{Cloud and Cluster Morphologies}
\label{sec:results-morph}

\begin{figure*}[htb]
	\centering
	\includegraphics[width=2.1\columnwidth]{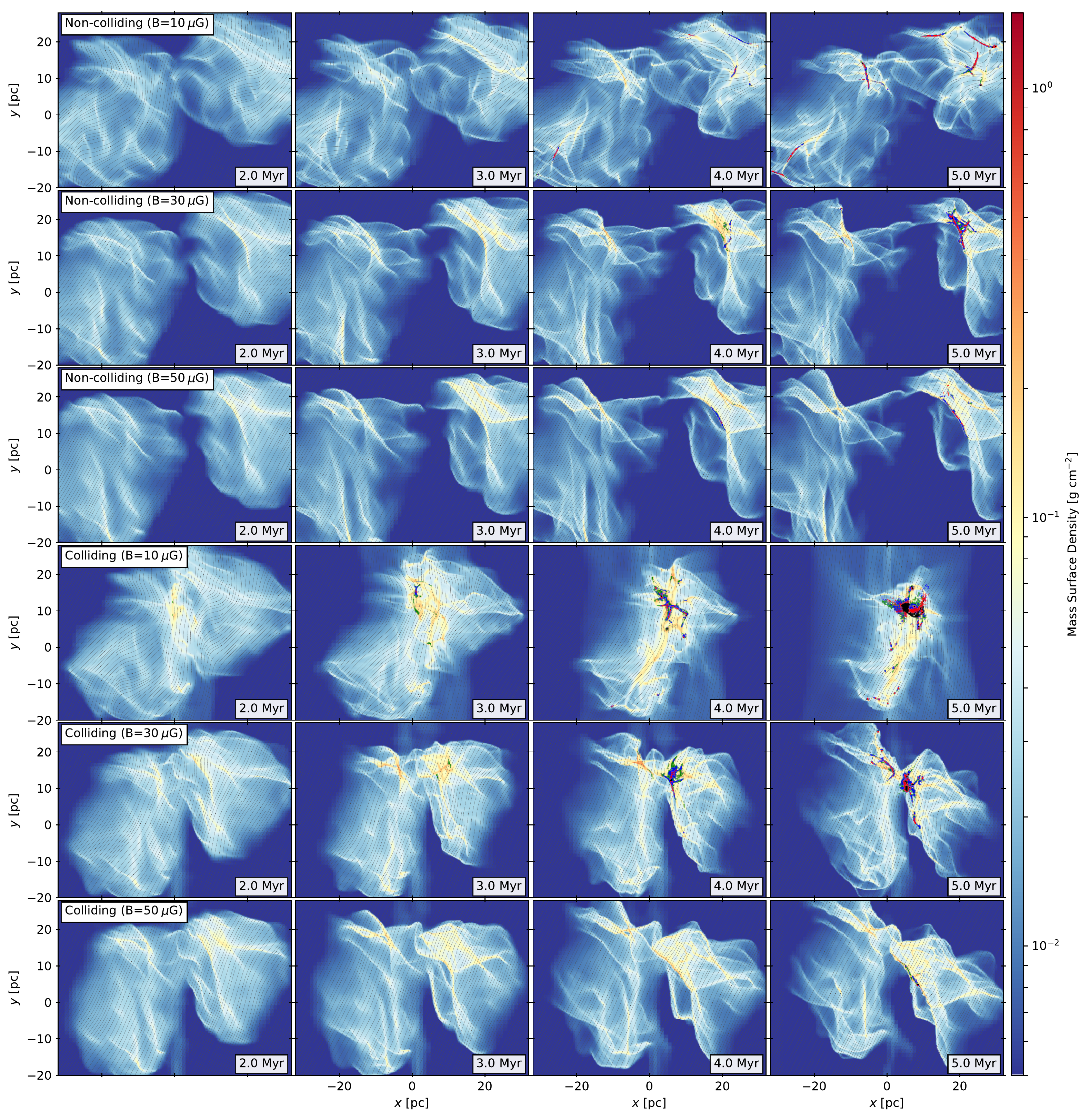}
	\caption{
		Time evolution of gas mass surface density, magnetic fields, and stellar distributions for all models with density regulated SF. 
		Snapshots at 2.0, 3.0, 4.0, and 5.0 Myr (left to right) are shown projected along the $z$-axis. The top three rows display the non-colliding runs with $B=$10, 30 and 50$\mu$G, respectively, while the bottom three rows display the colliding runs with $B=$10, 30 and 50$\mu$G, respectively. Density-weighted projected magnetic fields are shown as gray streamlines. The corresponding star particles are shown as black points. Stars created in simulations using the star formation models with $c_{1}=0.5 c_{\rm 1,fid}$, $1.0 c_{\rm 1,fid}$, and $2.0 c_{\rm 1,fid}$ are additionally plotted as blue, green, and red points, respectively. 
		\label{fig:sigma_SF}}
\end{figure*}

Gas and star particle distribution morphologies for the complete set
of simulations are shown in Figure~\ref{fig:sigma_SF}. The time
evolution of gas mass surface density and star particle distributions for
$B=$10, 30, and 50$\mu$G are displayed for each non-colliding and
colliding case where $c_{1}=c_{\rm 1,fid}$. Star positions from each
of the SF models are additionally plotted as separate colors. This was
chosen to more clearly compare the resulting star cluster
distribution, as the gas morphology does not differ significantly for
a given model based on our chosen SF model $c_{1}$
\citep[cf.,][]{Wu_ea_2017b}.

In general, the gas in non-colliding GMCs forms more dispersed
networks of filaments in contrast to the compressed structures arising
in colliding GMCs. The collision also enhances the mass surface
densities in both the filamentary gas and ambient material by factors
of a few or more. This enhancement occurs at earlier evolutionary
times relative to the non-colliding cases, due to the additional
accumulation of gas from the colliding flows.

The star clusters that form in each case are also morphologically
distinct, with the non-colliding clouds exhibiting more isolated
clusters elongated along the densest filaments while the colliding
clouds create one primary central cluster with smaller groupings
forming in dense knots within the central filamentary clump. The
density-regulated SF model forms clusters that span the smallest
extent, often at nodes where filaments intersect. As $c_{1}$ decreases
for the magnetically-regulated models, the magnetic criticality
threshold for star formation to occur is relaxed, increasing the
physical elongation of the star clusters and overall number of cluster
members along filamentary regions.

For non-colliding GMCs, the strength of the initial $B$-field moderately
influences the overall morphology. Self-gravity and turbulence form
distributed networks of filaments which gradually increase in mass
surface density. For weaker overall $B$-fields, the gas fragments and
collapses more readily, resulting in more numerous and physically
separated filaments. Stronger $B$-fields tend to create smoother, more
connected filamentary structures. In the weaker field case the
projected magnetic field has much greater curvature, while in the
stronger field case it is less perturbed from the initial
configuration. The star formation behavior is notably affected by the
$B$-field strength. With weaker fields, elongated star clusters are spread
throughout the greater GMC complex. For the 30 and 50$\mu$G cases,
star formation becomes concentrated in a single high density
region. The density-regulated SF model does not form stars in the
latter case.

The strength of the $B$-field significantly affects the evolution of the
colliding GMC models. In the weaker field case, both GMCs and $B$-fields
are strongly compressed in the central colliding region, forming a
large dense filament aligned perpendicular to the collision axis with
$B$-fields generally reoriented parallel to this filament at large
scales. For stronger field cases, the anisotropy of the magnetic
pressure causes the collision to become less direct. Gas is still
compressed in the colliding region, but the $B$-fields play a more
dominant role, preferentially guiding gas along the field lines. These
combined effects result in dense filaments that form within the
colliding region, which become qualitatively more perpendicular to the
mean field direction as the field strength increases. Star cluster
formation occurs within these dense regions. In our models, the
surrounding low density CNM contains higher magnetic pressure and
effectively creates a barrier inhibiting direct merging of gas between
the two GMCs, especially in the 50$\mu$G case. This may indicate that
GMC interactions in higher $B$-field environments are more indirect in
nature or may preferentially occur along field lines.

\subsection{Magnetic Field Orientations and Strengths}
\label{sec:results-Bfields}

The role of $B$-fields in star formation is closely tied with how they
influence molecular cloud gas evolution. In super-Alfv\'enic regimes
(i.e., $B$-fields are weak relative to kinetic motions), gas flows
dominate the morphology of the $B$-field, yet increase the field
strength along compressed regions. In sub-Alfv\'enic regimes (i.e.,
$B$-fields are strong relative to kinetic motions), the $B$-field
dictates the gas flow along field lines. Elucidating this mutual
connection has become an active field of research \citep[see, e.g.,
  reviews by][]{Hennebelle_Inutsuka_2019, Krumholz_Federrath_2019} and
is buttressed by expanding polarization capabilities of contemporary
observational facilities. Two major elements to consider are how the
direction of the $B$-field is correlated with gas structures and how
the strength of the $B$-field correlates with the density field.

\subsubsection{$B$-field versus Filament Relative Orientations}
\label{sec:results-Bfields-HRO}

\begin{figure*}[htb]
\centering
\includegraphics[width=1.8\columnwidth]{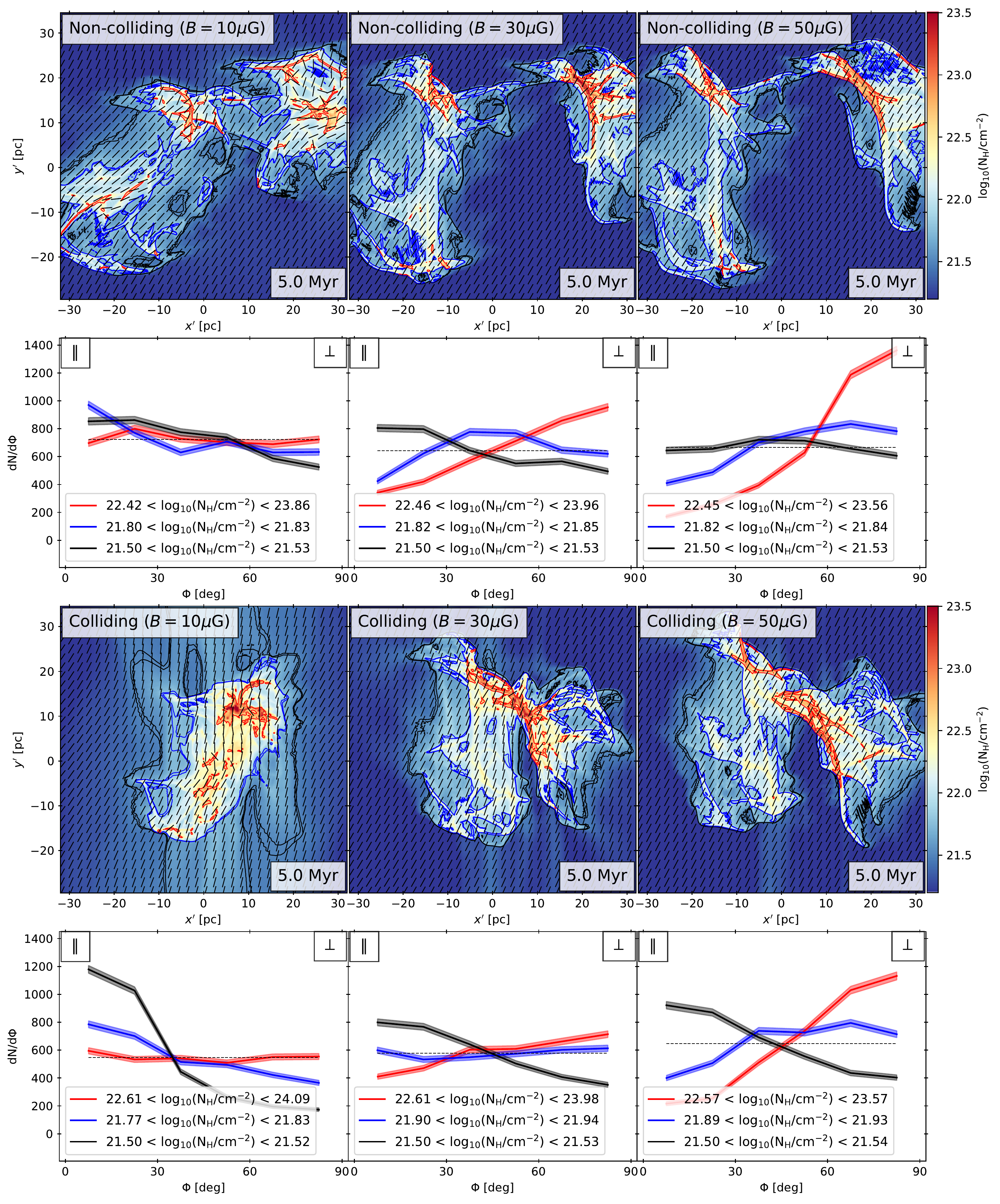}
\caption{
First and third rows: Column density maps with the normalized
plane-of-sky magnetic polarization field, $p$, represented with black
pseudo-vectors. The non-colliding cases for $B=10$, 30 and 50 $\mu$G
are shown in the first row, while the respective colliding cases are
shown in the third row. Gas from the density regulated SF models are
shown in each case at $t=5\:{\rm Myr}$. Second and fourth rows: Histogram of
Relative Orientations (HROs) are shown directly below their
corresponding column density maps. HROs compare the angle between $p$
vs. iso-$N_{\rm H}$ contours pixel-by-pixel. The projected map is
divided into 25 column density bins of equal pixel count. HROs for the
lowest (1st bin; black), middle (12th bin; blue), and highest (25th
bin; red) $N_{\rm H}$ are shown, using angle bins of $15\degree$. The
histogram color corresponds to the regions bounded by low (black),
intermediate (blue), and high (red) value iso-$N_{\rm H}$ contours as
shown in the respective projection map. The thickness of the data
points indicates the variance within each bin. Histograms with peaks
at $0\degree$ correspond to $p$ predominantly aligned with iso-$N_{\rm
  H}$ contours (i.e., $B$-fields parallel to filaments), while peaks
at $90\degree$ correspond to $p$ predominantly perpendicular to
iso-$N_{\rm H}$ contours (i.e., $B$-fields perpendicular to
filaments).
\label{fig:HRO}}
\end{figure*}

The formation and evolution of filamentary structures in molecular
clouds may be strongly affected by the orientation of the $B$-field.
\citet{Li_ea_2013} observed a bimodal distribution of preferentially
parallel and perpendicular orientations between filamentary Gould Belt
clouds and their encompassing $B$-fields. They concluded that dynamically
important $B$-fields must be present, which both guide the gravitational
contractions (perpendicular orientations) and channel turbulence
(parallel orientations). If the primary filament formation mechanism
were instead super-Alfv\'enic turbulence there should be no
preferential alignment.

\citet{Soler_ea_2013} developed a pixel-by-pixel approach to quantify
the degree of alignment of magnetic fields with respect to filamentary
structures defined by column density gradients. This statistical
method, the Histogram of Relative Orientations (HRO), measures the
relative orientations for a given (column) density and has since been
widely used in both polarization observations
\citep[e.g.,][]{PlanckXXXV_2016} and numerical simulations
\citep[e.g.,][]{PlanckXX_2015,Chen_ea_2016,Wu_ea_2017a}.

For simulations described in the current work, we apply an HRO
formulation where
\begin{equation}
    \Phi = \left| \arctan \left( \frac{\nabla N_{\rm H} \cdot
  \bm{p}}{|\nabla N_{\rm H} \times \bm{p}|} \right) \right|
\end{equation}
is the magnitude of the angle between $N_{\rm H}$ iso-contours 
(orthogonal to $\nabla N_{\rm H}$) and $\bm{p}$, the simulated 
polarization pseudo-vectors. $\bm{p}$ is defined as
\begin{equation}
\bm{p} = (p \sin \chi)\hat{\bm{x}} + (p \cos \chi)\hat{\bm{y}}
\end{equation}
where $p=0.1$ is a set constant polarization fraction and $\chi$ is the
angle in the plane-of-sky derived from the Stokes parameters \citep[see, e.g.,][]{Wu_ea_2017a}:
\begin{equation}
q = \int n \frac{B_{y}^{2}-B_{x}^{2}}{B^{2}} ds
\end{equation}
\begin{equation}
u = \int n \frac{2B_{x}B_{y}}{B^{2}} ds
\end{equation}
\begin{equation}
\chi = \frac{1}{2} \arctan2(u,q)
\end{equation}

The relative orientation angle $\Phi$ is then
calculated pixel-by-pixel in a column density map. 
The lowest column density pixels ($N_{\rm H} < 21.5\:{\rm cm^{-2}}$) 
are ignored and 
a gradient threshold for $\nabla N_{\rm H}>0.25\overline{N_{\rm H}}$
is applied \citep[similar to, e.g.,][]{PlanckXXXV_2016}, to 
better separate GMC material from diffuse background fields.
The remaining pixels are then separated into 25 bins of equal count
based on $N_{\rm H}$. Note that we assume $n_{\rm He}=0.1\:n_{\rm H}$,
yielding a mass of $2.34\times 10^{-24}\:{\rm g}$ per H.

Figure~\ref{fig:HRO} shows column density maps for the non-colliding
and colliding models with $B=10$, 30, 50 $\mu$G, respectively, for the
density-regulated model. Their respective HROs are also shown, for
low, medium, and high column density bins. Here, $\phi$ represents the
smaller angle between the polarization-inferred magnetic field and
$N_{\rm H}$ iso-contours. Thus, histograms peaking at $\phi=0\degree$
indicate inferred $B$-fields preferentially aligned parallel to
filamentary structures, while peaks at $\phi=90\degree$ indicate a
preferentially perpendicular alignment.

Gas in the non-colliding, $B=10\:{\rm \mu G}$ model is slightly preferentially
aligned parallel to filaments at high, medium, and low column
densities. At $30\:{\rm \mu G}$, the low-$N_{\rm H}$ gas retains effectively
the same orientation behavior, while the mid- and high-$N_{\rm H}$
bins show an increasing trend toward perpendicular alignment. This
trend increases further in the $50\:{\rm \mu G}$ case, where the low-$N_{\rm
  H}$ gas has random alignment, mid-$N_{\rm H}$ gas has slightly
perpendicular alignment, and high-$N_{\rm H}$ gas is strongly
perpendicularly aligned with the $B$-fields. 

The colliding GMC models exhibit quite different relative
orientations. The low-$N_{\rm H}$ gas in the $B=10 \mu$G model is
strongly aligned parallel to the $B$-fields, due to the large-scale
flows reorienting the fields perpendicular to the velocities. The
degree of alignment decreases for increasing $N_{\rm H}$ bins, with
essentially no preferential alignment for the highest-$N_{\rm H}$
regions. Similar to the non-colliding cases, the alignment of
filamentary structures shifts towards more perpendicular relative
orientations in higher $B$ environments. However, the lowest-$N_{\rm
  H}$ structures remain fairly preferentially oriented parallel to the
$B$-fields, due to effects of the colliding flows on the ambient gas.

Observationally, HROs from active star-forming regions typically
reveal polarization vectors oriented in a preferentially parallel
alignment with low-$N_{\rm H}$ gas structures. For high-$N_{\rm H}$
gas structures, there tends to instead be a random or preferentially 
perpendicular alignment, 
\edit1{though high variation in morphology is observed \citep[see, e.g.,][]{PlanckXXXV_2016}}. 
From our simulations, distinctive signs of a
collision include a strongly parallel alignment of low-density gas,
though lower $B$-field non-colliding models also exhibit this behavior
but to a lesser extent. In both types of models, the strength of the 
$B$-field has the greatest impact on forming perpendicular alignments 
with high-$N_{\rm H}$ gas in the HRO.
Overall, the degree of separation for low- and high-$N_{\rm H}$ gas in
HROs may act as a supplementary tool to investigate both the large 
scale dynamics and magnetic field strength of star-forming regions.

\begin{figure*}[htb]
\centering
\includegraphics[width=2.1\columnwidth]{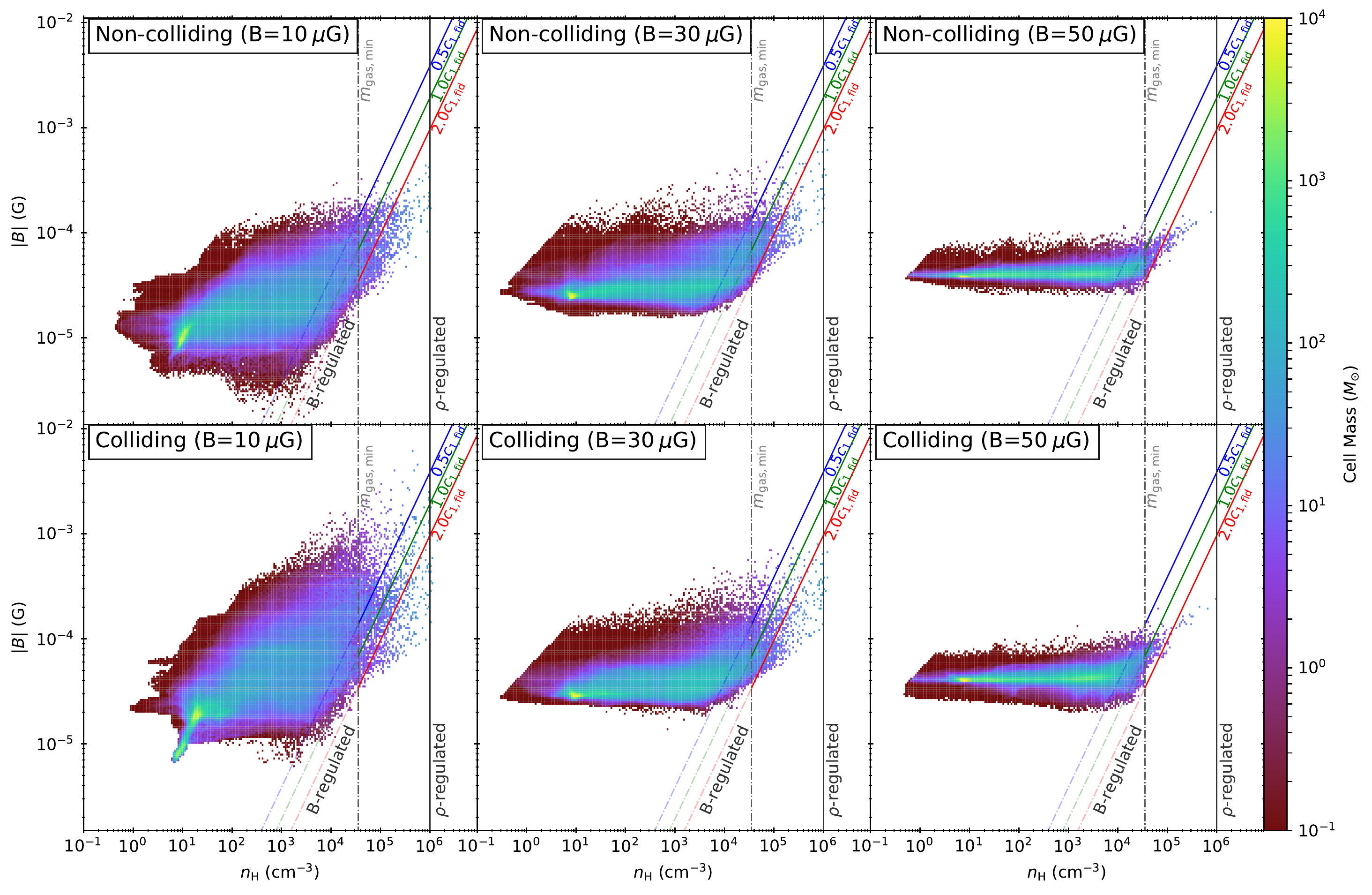}
\caption{
Phase plots of $B$ versus $n_{\rm H}$ for the non-colliding (top row)
and colliding (bottom row) cases at 5~Myr.  The first, second, and
third columns show the $B=10$, 30 and 50 $\mu$G models, respectively,
for cell mass distributions using the density regulated SF model. The
black vertical line represents the corresponding $n_{\rm H,sf}$
threshold. The blue, green, and red lines represent the various
mass-to-flux thresholds for the magnetically-regulated star formation
routine. Star particle formation proceeds for densities above these
respective limits.  The black dash-dot line shows the effective
minimum gas density threshold for star formation due to the 50\% mass
limitation of $m_{\rm \star,min}$ assumed in the sub-grid routines.
The Alfv\'en limiter described in \S~\ref{sec:methods-code} imposes an
effective density floor that can be seen in certain models in the low
density, higher field strength regimes.
\label{fig:SF_phase}}
\end{figure*}

\subsubsection{$B$-field Strength vs. Density}
\label{sec:results-Bfields-Bvsn}

Zeeman observations estimating the magnetic field strength with
respect to the gas density in the ISM have found a relatively constant
value of $B=10\mu$G for $n\lesssim 300\:{\rm cm^{-3}}$ and an
approximately power law relation, $B\propto n_{\rm H}^\kappa$, above
this threshold \citep{Crutcher_ea_2010}. The precise value of $\kappa$
has been debated, though indices near $\simeq 0.65$ or 0.5 are
estimated, where the former would arise in idealized spherical
contraction and the latter from non-isotropic
contraction. \citet{Mocz_ea_2017} found $\kappa \simeq 0.65$ in their
simulations of super-Alfv\'enic turbulence and $\kappa \simeq 0.5$ for
sub-Alfv\'enic conditions, in agreement with expected gas contraction
behavior under weak and strong $B$-fields, respectively.

We investigate the $B$ versus $n_{\rm H}$ relation in our suite of GMC
simulations, with the results shown in
Figure~\ref{fig:SF_phase}. These phase plots show the total cell mass
at each $B$ and $n_{\rm H}$ for the various models, as well as the
minimum mass and mass-to-flux thresholds used in the star formation
routines.

For the non-colliding case at $B=10\mu$G, a fairly wide range of
magnetic field strengths exists for a given density. The peak
distribution in gas mass can be attributed to the initially uniform
ambient medium. The remaining gas mass generally follows a positive
correlation between $B$ and $n_{\rm H}$. In the $30\mu$G model, the
overall spread in $B$ is narrower, with an increased average $B$
but similar maximum value. In the $50\mu$G model, the spread in $B$
decreases even further, and the maximum values for both $B$ and
$n_{\rm H}$ reached in the simulation are the lowest of the three. As
the initial $B$-field strength increases, the overall collapse of the
GMCs is strongly inhibited, leading to a more uniform distribution of
final $B$-field strength.

The colliding GMC models follow a similar trend where higher global
field strengths result in narrower overall distributions in
$B$. However, the collisions impart moderately wider spreads in both
density and $B$-field strength relative to their non-colliding
counterparts. The highest values in both density and $B$-field
strength are achieved in the colliding $B=10\mu$G case.

In both scenarios, the effective $\kappa$ index in the $B$ versus
$n_{\rm H}$ relation decreases as the average initial $B$-field
strength increases. This is consistent with the results found when
comparing gas contraction in sub- and super-Alfv\'enic environments.

Attributes of the star formation routines can be gleaned from these
plots as well. Thresholds for the density- and magnetically-regulated
models are shown, above which gas will be converted into stars
following the methods described in \S\ref{sec:methods-SF}. For the
density-regulated model, this is a simple density threshold. For the
magnetically-regulated models, both the mass-to-flux ratios above the
respective $c_{1}$ level and the threshold density dictated by $m_{\rm
  \star, min}$ must be satisfied. The $c_{1}=0.5 c_{\rm 1, fid}$
routines yield the greatest total gas mass to be converted into stars
in each of the GMC evolution models, resulting in the widest
proliferation of star formation as seen in
Figure~\ref{fig:sigma_SF}. Likewise, models with $c_{1}=2 c_{\rm 1,
  fid}$ generally have less available gas mass and less star
formation. However, that which does occur tends to do so in higher
density cells at higher mass-to-flux ratios. This leads to different
stellar and star-forming gas properties, as detailed in later sections.

\subsection{PDFs of Mass Surface Density}
\label{sec:results-PDFs}

Probability distribution functions (PDFs) are a useful statistical tool for connecting gas distributions with the physical mechanisms that shape them. Certain properties of PDFs have been shown in simulations to reveal regimes dominated by turbulence or self-gravity as well as being sensitive to pressure from, e.g., shocks and magnetic fields \citep[][]{Vazquez-Semadeni_1994,Padoan_ea_1997a,Kritsuk_ea_2007,Federrath_ea_2008,Price_2012,Collins_ea_2012,Burkhart_ea_2015}. Observationally, PDFs of mass surface density $\Sigma$ (also $N_{\rm H}$ or $A_{\rm V}$) provide an important link to simulations and have been used to infer underlying physical characteristics of molecular clouds as well as IRDCs \citep[e.g.][]{Kainulainen_ea_2009,Kainulainen_Tan_2013,Butler_ea_2014}.

An area-weighted $\Sigma$-PDF, $p_{A}$, can often be well fit by a lognormal of the form
\begin{equation}
    p_{A}(\ln \Sigma') = \frac{1}{(2 \pi)^{1/2}\sigma_{\ln \Sigma'}} \left( \exp{-\frac{(\ln \Sigma'-\overline{\ln \Sigma'})^{2}}{2 \sigma^{2}_{\ln \Sigma'}}} \right)
\end{equation}
where $\Sigma'\equiv \Sigma/\overline{\Sigma_{\rm PDF}}$ is the mean-normalized $\Sigma$ and $\sigma_{\ln \Sigma}$ is the lognormal width, which increases with higher turbulent Mach numbers. Power-law tails often form at high-$\Sigma$, indicative of the degree of gravitational collapse and correlated with the efficiency of star formation. 

\begin{figure*}
\centering
\includegraphics[width=1\columnwidth]{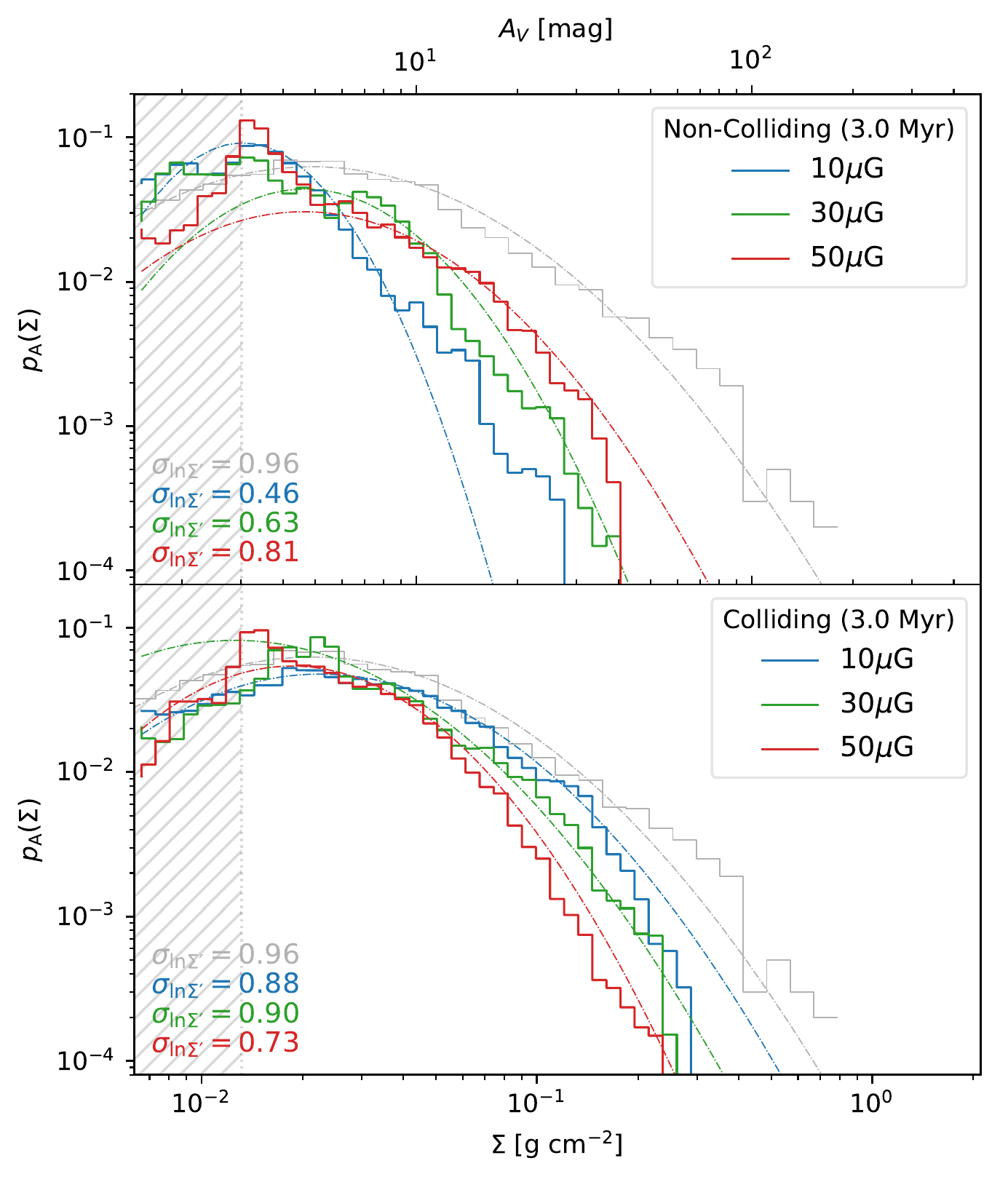}
\includegraphics[width=1\columnwidth]{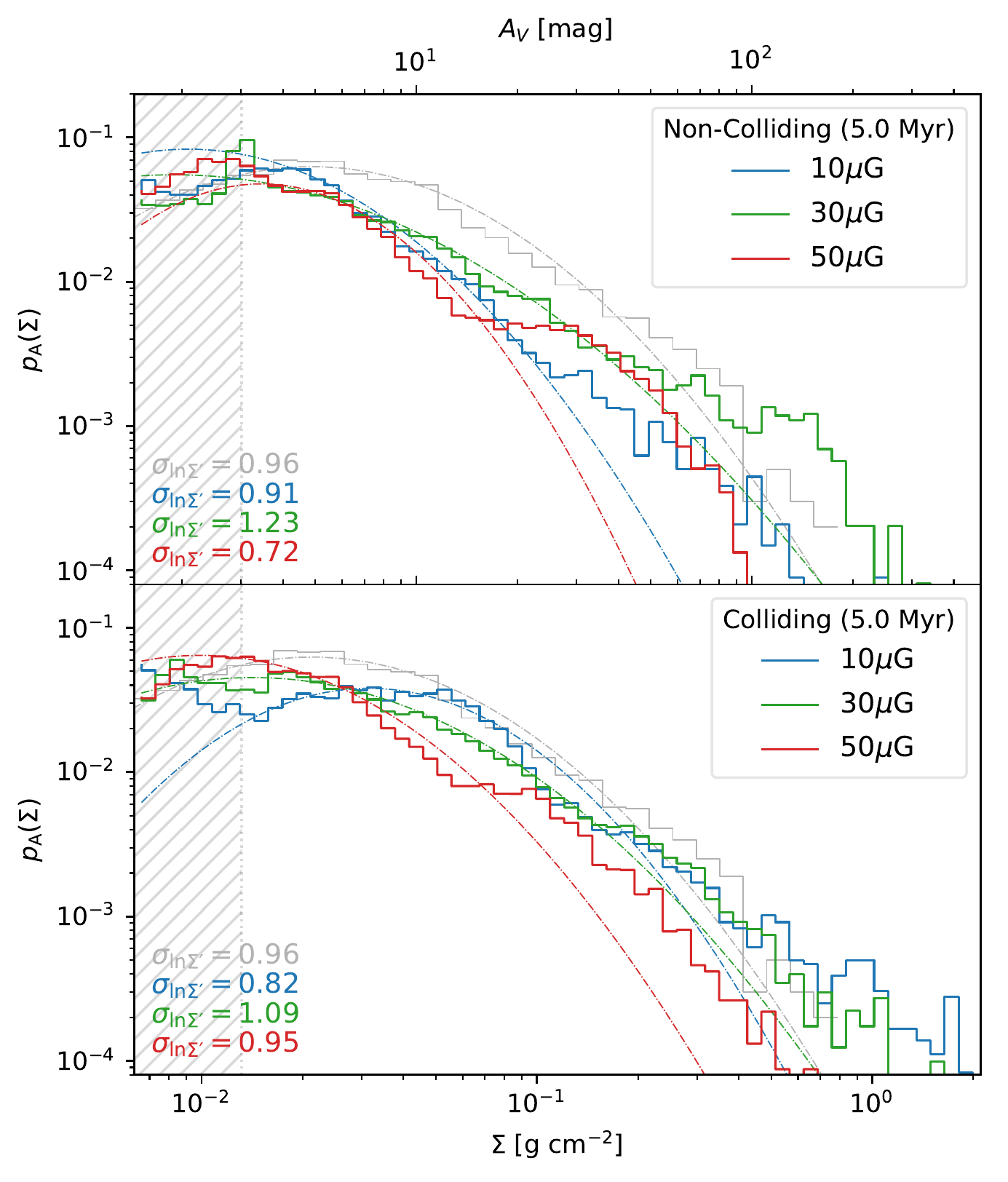}
\figcaption{
Area-weighted $\Sigma$-PDFs of $(32\:{\rm pc})^3$ regions from non-colliding
(top panels) and colliding (bottom panels) simulations for $t$=3.0 
(left column) and 5.0 Myr (right column). For each case, gas distributions
are shown for the $B=10$, 30, and 50\:$\mu$G models with density-regulated 
star formation. 
In the same color, lognormal fits for each case are plotted as dash-dotted 
lines and the corresponding width $\sigma_{\ln \Sigma}$ is shown.
Also displayed in each panel 
\edit1{in gray} 
is the observed $\Sigma$-PDF for the massive 
IRDC from \citet{Lim_ea_2016}. The shaded region denotes areas of 
$A_{V}<3\:{\rm mag}$ corresponding to the lowest closed contour level of
the observational data: distributions are only fit above this level.
\label{fig:PDFstime}}
\end{figure*}

Figure~\ref{fig:PDFstime} shows area-weighted $\Sigma$-PDFs at different 
evolutionary times in non-colliding and colliding simulations where $B=10$, 
30, and 50\:$\mu$G. These $(32\:{\rm pc})^3$ regions are centered on the 
gas density maximum and projected along the $z'$-direction. 

At $t=3\:{\rm Myr}$, gas distributions in the non-colliding models peak at 
approximately $0.2\:{\rm g~cm^{-2}}$, with weaker initial $B$-fields 
corresponding to a relatively lower distribution of gas at high $\Sigma$ 
values. Colliding models reach higher values of $\Sigma$ by the same 
evolutionary time, with weaker $B$-field models experiencing higher 
relative increases. This results in a slight reversal of the trend, where 
collisions in weaker $B$-fields create slightly greater concentrations of 
material at higher $\Sigma$. The $\sigma_{\ln\Sigma'}$ values for the 
colliding models are greater, on average, 
and, especially for the more evolved case at 5~Myr, are a closer match to 
the $\Sigma$ distribution observed in the massive IRDC by \citet{Lim_ea_2016}.

The gas develops much higher distributions of gas at high-$\Sigma$ in all 
cases by $t=5\:{\rm Myr}$. Gas exceeds $0.5\:{\rm g~cm^{-2}}$ in both the 
non-colliding and colliding cases, and develops distinct high-$\Sigma$ 
material not well-fit by the lognormals. The $10\:\mu$G colliding run exceeds 
$1.0\:{\rm g~cm^{-2}}$, while the maximum surface density decreases as
initial $B$-field strengths increase. The non-colliding models do not appear 
to exhibit strong trends based on $B$-field strength.
Here, $\sigma_{\ln\Sigma'}$ reaches higher values in all cases, with the 
colliding models generally exhibiting larger widths.  

Overall, the $\Sigma$-PDFs reveal that generally greater amounts of 
high-$\Sigma$ gas form in collisions, while the strength of the $B$-field
does not appear to play a large role in shaping the resulting PDFs. 
The greatest changes in surface density distributions occur with 
time-evolution, pulling higher concentrations of gas toward the 
high-$\Sigma$ end and  signifying the dominant role of gravity at 
later stages.

\begin{figure*}[htb]
\centering
\includegraphics[width=2.1\columnwidth]{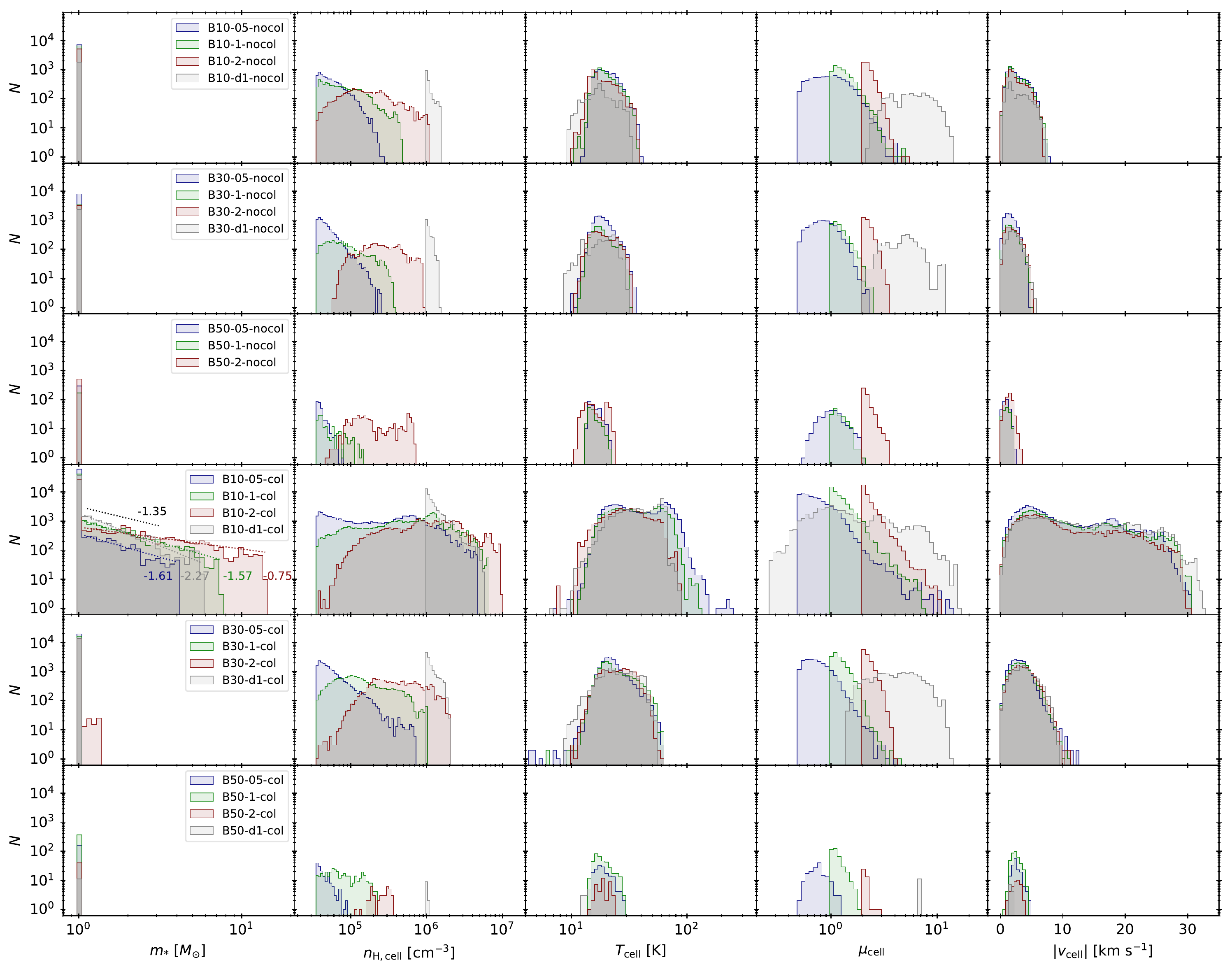}
\caption{
Masses of star particles and properties of star-forming gas cells at
the time of star particle creation.  (Left to right \edit1{columns}:) Cumulative
histograms for each simulation up to $t=5$ Myr of star particle mass
(bin width is 0.03 dex), cell Hydrogen number density (bin width is
0.03 dex), cell gas temperature (bin width is 0.03 dex), cell normalized
mass-to-flux ratio (bin width is 0.03 dex), and cell velocity magnitude
(bin width is 0.45 km/s) are plotted.  The non-colliding (top three
rows) and colliding (bottom three rows) cases are shown, for $B=10$,
30, and 50 $\mu$G, respectively. For each case, the results from the
density regulated SF routine is shown in gray, while \edit1{results from} the magnetically
regulated SF routines with $c_{1}=0.5, 1$ and $2 c_{\rm 1,fid}$ are
displayed in blue, green, and red, respectively. 
\edit1{Stellar mass histograms for the $B=10\mu$G colliding cases (row 4)} 
also include power law fits for the
distribution of stars with $m_{\star}>1~M_{\odot}$. The power law
indices are displayed, as well as a reference -1.35 (Salpeter) index.
\label{fig:SFcells_hist}}
\end{figure*}

\subsection{Properties of Star-Forming Gas}
\label{sec:results-SFcells}

We investigate stellar masses and their natal gas properties at the
time of star birth, which can be considered to be an approximate
representation of the properties of pre-stellar cores in these models,
with the caveat that the finest grid resolution is a relatively large
0.125~pc. Figure~\ref{fig:SFcells_hist} plots the cumulative
histograms for each model, through 5~Myr, of initial star particle
masses and progenitor cell gas densities, temperatures, mass-to-flux
ratios, and velocity magnitudes in the global center-of-mass frame.

In the non-colliding models, all stars form at the threshold mass of
$m_{\rm \star, min}=1~M_{\odot}$. This indicates that star formation
occurs solely in the stochastic regime of the models, i.e., all stars
would have formed with $m_{\star} \leq 1~M_{\odot}$. A total number
approaching $10^{4}$ are created in the $10\mu$G models, with stronger
$B$-fields leading to more than a factor of 10 lower overall star
formation. 
\edit1{While higher mass stars may be expected to form from a typical
$\sim10^{5}\:M_{\odot}$ GMC in the Milky Way, these models inform us 
that no region within our simulated non-colliding GMCs reaches a high
enough mass accretion rate by $t=5\:{\rm Myr}$, 
(i.e., $\dot{m}_\star \Delta t \le m_{\rm \star,min}$ 
as described in \S\:2.3).}

For star-forming cells in the non-colliding models, the ranges of
densities, temperatures, mass-to-flux ratios, and velocities also
decrease for higher $B$-field models. As outlined in
\S\ref{sec:results-Bfields-Bvsn}, the stronger $B$-fields tend to
constrain the overall variance in gas density, which is generally
correlated with the other properties. For the density-regulated star
formation model, no stars are created in the $50\mu$G case, i.e., no
cells achieve the threshold density of $n_{\rm H}=10^6\:{\rm
  cm}^{-3}$. For the other cases, star-forming cells lie near the
density threshold, while in general, $\mu_{\rm cell} >2$. The
magnetically-regulated models form stars in all cases. The $c_{1}$
critical mass-to-flux ratio strongly affects the cell density
distribution, but is essentially independent from the cell temperature
and velocity.

Star formation occurs in a markedly different manner for the colliding
GMC models. In all $10\mu$G cases, a significant number of stars form
outside the stochastic regime, representing the occurrence of
higher-mass star formation. Following Equation~\ref{eq:SFRmdot}, this
requires the natal gas cells to accumulate large masses within short
timescales. The maximum stellar mass ranges from 4~$M_{\odot}$ in the
$0.5 c_{\rm 1, fid}$ model to 14~$M_{\odot}$ in the $2 c_{\rm 1, fid}$
model. While a fewer total number of stars form with the $2 c_{\rm 1,
  fid}$ model, higher stellar masses are achieved. This can be
understood by the higher local mass-to-flux ratio required to initiate
star formation, thus enabling cells to reach higher densities just
prior to birth of the star.

We fit power laws to these higher-mass stellar distributions, finding
indices of -2.27 for the density-regulated model and -1.61, -1.57 and
-0.75 for $c_{1}=0.5, 1$ and $2 c_{\rm 1,fid}$, respectively. The
density-regulated model is steeper than the reference -1.35 index,
often adopted for the observed IMF \citep[see, e.g.,
][]{Salpeter_1955}, while the magnetically-regulated models have a
range of values that are closer to the Salpeter index, with the higher
mass-to-flux ratio threshold leading to the most top-heavy IMF.
While other factors that are not yet included, especially protostellar
outflows and other forms of feedback may affect the IMF, in the
context of our star formation sub-grid models, we see that collisions
of relatively weakly magnetized GMCs enable the formation of more
massive stars by allowing the rapid accumulation of mass into cells at
rates that are faster than can be removed by formation of low-mass
stars.
\edit1{This variation may have some connection to the non-Salpeter,
top-heavy core mass functions (CMF) observed in massive protoclusters
and IRDCs \citep{Motte_ea_2018,LiuM_ea_2018}, 
which will be further explored with the ALMA-IMF large program (Motte et al., in prep).}

In the higher magnetization colliding models, only the $2 c_{\rm 1,
  fid}$ SF routine for $30\mu$G produces stars with masses above the
stochastic regime. Otherwise, they follow a similar trend as the
non-colliding cases, with stronger $B$-fields inhibiting overall star
formation.
\edit1{Tentatively, these results may suggest that a preferentially 
top-heavy IMF may form in more magnetically supercritical collisions. 
However, we note that our SF models are ultimately simple approximations 
and thus not necessarily representative of a realistic IMF.}

The cells in which stars form exhibit much greater variance in gas
properties due to the collisions. In the $10\mu$G cases, the
star-forming cells reach densities up to 10 times greater than their
non-colliding counterparts. Mean temperatures, mass-to-flux ratios,
and velocity magnitudes also increase, as do their variances. The
strength of the initial $B$-field plays a large role in the resulting
properties found in star-forming gas, with higher $B$-fields
constraining the range of the gas properties shown. Differences
arising from the collision become much less pronounced in the presence
of stronger initial $B$-fields.

The distributions of velocity are closely correlated with the velocity
dispersions of the primary clusters (\S \ref{sec:results-clustering})
in the colliding models. This indicates that a significant fraction of
the stars form in the potential of the primary cluster. The weaker
correlations for the non-colliding cases are in line with their more
dispersed cluster formation, where no single primary cluster dominates
the stellar dynamics.

\begin{figure*}[htb]
\centering
\includegraphics[width=2.1\columnwidth]{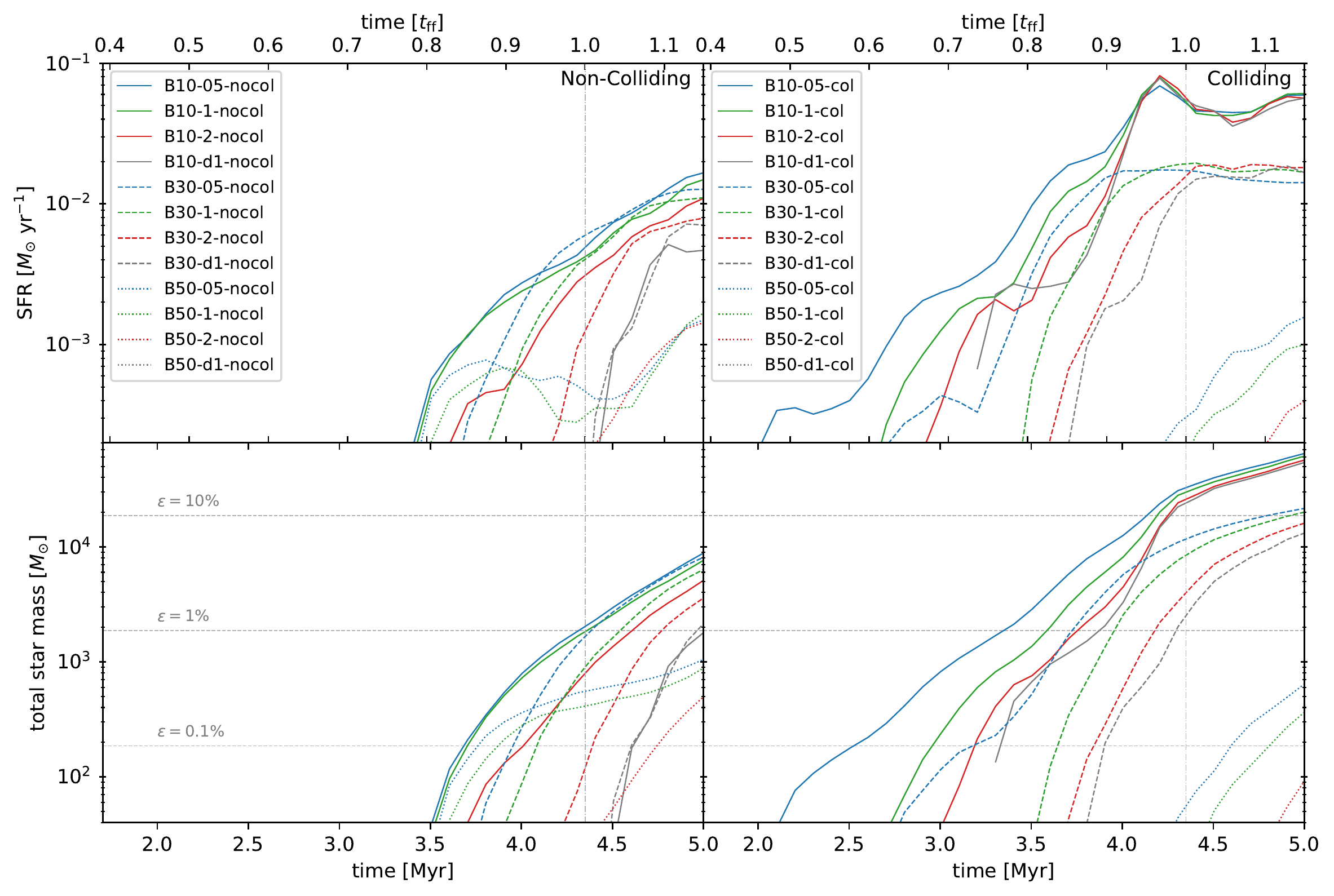}
\caption{
{\it Top row:} Star formation rates as a function of time.  The
non-colliding cases are shown in the left panels, while the colliding
models are shown in the right panels. Simulations with $B=10$, 30 and
$50\mu$G are shown with solid, dashed and dotted lines,
respectively. The density-regulated SF routine is shown in gray, while
the magnetically-regulated SF routines with $c_{1}=0.5, 1$ and $2
c_{\rm 1,fid}$ are displayed in blue, green and red, respectively. The
time is shown in both simulation time (bottom axis) and free-fall
times of the initial GMCs (top axis).  The vertical dotted line
indicates one initial free-fall time, i.e., $t_{\rm ff}=4.35$~Myr.
{\it Bottom row:} The cumulative mass of created stars over time. The
star formation efficiency, with reference values shown for
$\epsilon$=0.1\%, 1\% and 10\%, is normalized relative to the total
initial mass of the two GMCs ($1.86\times 10^{5}\:M_{\odot}$).
\label{fig:SFR}}
\end{figure*}

\subsection{Star Formation Rates and Efficiencies}
\label{sec:results-SFR}

We next investigate how GMC collisions and $B$-field strength affect the
overall SFR and SFE.  Figure~\ref{fig:SFR} shows the SFR and
cumulative mass of formed stars for each model as a function of
simulation time and freefall time of the initial GMCs. The value for
SFR at a given time is calculated as the time derivative of the total
star mass. Levels for SFE are shown as the total star mass normalized
by the combined gas mass of the original two GMCs.

In the non-colliding models, star formation is initiated around
$t=3.5~$Myr at the earliest, and then showing a general increase in
SFR over the course of the simulation. The density-regulated models
begin forming stars only after 1~$t_{\rm ff}$, with almost identical
behavior for the 10 and $30\mu$G cases, while no stars form in the
$50\mu$G case. In the magnetically-regulated models, reduced $c_{\rm
  1}$ thresholds lead to earlier star formation, higher SFRs, and
higher SFEs. For the $10\mu$G case, SFRs reach $\sim 5\times
10^{-3}\:M_{\odot}\: {\rm yr^{-1}}$ with a SFE of 1\% by $t_{\rm
  ff}$. Stronger field strengths lead to lower overall SFRs and SFEs.

Star formation commences in the colliding GMCs at roughly 2.5, 3.2,
and 4.5~Myr for $B=10$, 30 and $50\mu$G, respectively. Within each
$B$-field case, as the SF model changes from $c_{\rm 1}$=0.5, 1.0, 2.0,
to density-regulated, star formation begins later, SFRs are lower, and
SFEs are lower. In all cases, SFEs increase monotonically, while SFRs
increase and then begin to level off. The SFRs in the 10 and $30\mu$G
cases reach approximately $7\times10^{-2}$ and
$2\times10^{-2}~M_{\odot}\: {\rm yr^{-1}}$, respectively. In the
$50\mu$G case, stars only
begin to form at $t=5~$Myr for the density-regulated model. Overall, star
formation efficiencies per freefall time are approximately 15\%, 2-5\%
and 0\%, respectively as magnetic field strength increases.

Relative to the non-colliding models, the collision triggers earlier
star formation by 1~Myr and enhances SFRs and SFEs by over a factor of
10 in the weaker $B$-field cases. However, as $B$ increases, the
enhancement of star formation activity due to collisions is less
prominent, and, in the case of $50\mu$G, actually inhibited. This
behavior can be attributed to the higher magnetic pressure especially
in the bounding atomic regions that lead to a dampening of the
collision. The collision cannot efficiently accumulate gas into dense
clumps as the magnetic pressure acts to inhibit the flow of gas toward
any converging region. These results may be applicable to regions with
high densities, magnetic field strengths, and turbulence, yet
relatively low SFRs, such as the Central Molecular Zone \citep[e.g.,
][]{Kruijssen_ea_2014}, including the ``Brick'' IRDC \citep[e.g.,
][]{Henshaw_ea_2019}. However, equivalent simulations for these higher
density conditions would need to be carried out to confirm this
hypothesis.

The near-convergence of the different star formation models at lower
$B$-field strengths indicate that the SFR and SFE are not
significantly limited by the density and mass-to-flux thresholds set
by our simulations (see also Paper III). Instead, they seem to be
determined by the creation of larger structures that contain gas with
values greater than the thresholds, which are then converted
efficiently into stars even with the $\epsilon_{\rm ff}=0.02$ rate
within the star-forming cells. As the $B$-field strength increases,
variations among the SF models lead to divergence of the SFRs and
SFEs.

Note that in each of these star formation models, key stellar feedback
processes such as protostellar outflows, ionization, winds, and
radiation pressure have not yet been implemented. While the presented
simulations may approximate the initial onset of star formation, the
aforementioned feedback mechanisms will likely reduce SFRs at later
times.

\subsection{Spatial Clustering and Dynamics of Primary Cluster}
\label{sec:results-clustering}

We investigate how the spatial clustering of stars changes with
magnetic field strength and in the presence of collisions.  To analyze
the overall clustering behavior over the entire GMC complex, we use
the \textit{minimum spanning tree} (MST) which can determine the
degree of centrally concentrated clustering versus multi-scale
clustering.  The substructure within the most massive cluster is
explored using the \textit{angular dispersion parameter} (ADP), a
metric that is sensitive to azimuthal variations for chosen
radii. Additionally, we calculate the virial parameter to estimate the
dynamical state of the primary cluster.

The primary star cluster is found via the popular data clustering
algorithm, \textit{density-based spatial clustering of applications
  with noise} (DBSCAN) \citep{Ester_ea_1996}. DBSCAN defines clusters
based on groupings of points with many nearby neighbors and ignores
outliers with neighbors that are too far away. Here, we set the points
to be our projected star particle locations and set the recovered
cluster with the largest population as our primary cluster. To find
the center of this cluster, we follow the iterative method introduced
in Paper III. First, the median position of the set of cluster members
is used as an initial guess. Then, we center on this position a
circular aperture with an initial radius of 0.4\:pc, and determine a
new center based on the center of mass using stars only within this
aperture. We repeat this process while iteratively halving the
aperture radius until it reaches 0.1\:pc. This final defined center is
used in the subsequent analysis.

\subsubsection{Minimum Spanning Tree}

The MST is a graph theory technique which seeks to minimize the
lengths of the ``edges'' that connect all ``vertices'' of a connected,
undirected graph. It was first developed for astrophysical
applications by \citet{Barrow_ea_1985}, and enables the quantitative
study of hierarchical substructure of stellar distributions by setting
the edge weights to be projected euclidean distances between
individual stars.

\citet{Cartwright_Whitworth_2004} introduced the $\mathcal{Q}$
parameter to measure the degree of radial concentration in
clustering. This is given by
\begin{equation}
    \mathcal{Q} = \frac{\overline{m}}{\overline{s}}
\end{equation}
where $\overline{m}$ is the normalized mean edge length and
$\overline{s}$ is the normalized correlation length, given by: 

\begin{equation}
    \overline{m} = \sum^{N_{\star}-1}_{i=1} \frac{e_{i}}{\sqrt{N_{\star}A}}
\end{equation}
and
\begin{equation}
    \overline{s} = \frac{\overline{d}}{R_{\rm cl}},
\end{equation}
respectively. $N_{\star}$ is the total number of star particles, 
$e_{i}$ is the length of each edge (of which there are $N_{\star}$-1),
and $A=\pi R_{\rm cl}$ is the cluster area. $R_{\rm cl}$ is distance from 
the mean star positions to the farthest star, and $\overline{d}$ is the 
mean pairwise separation distance between the stars. 
These are discussed in more detail in \citet{Wu_ea_2017b}
\footnote{Note that in that work, there are errors in Equation 10, 
which should instead show the reciprocal, and Equation 12, 
which contained a superfluous factor of $(N_{\star}-1)^{-0.5}$. 
However, these typos do not affect the calculations and figures 
of that paper.}.

A threshold value $\mathcal{Q}_{0}=0.8$ separates star clusters in
more centrally condensed associations ($\mathcal{Q}>\mathcal{Q}_{0}$)
and those in more substructured, multi-scale associations
($\mathcal{Q}<\mathcal{Q}_{0}$).

\begin{figure}
  \centering
  \includegraphics[width=1\columnwidth]{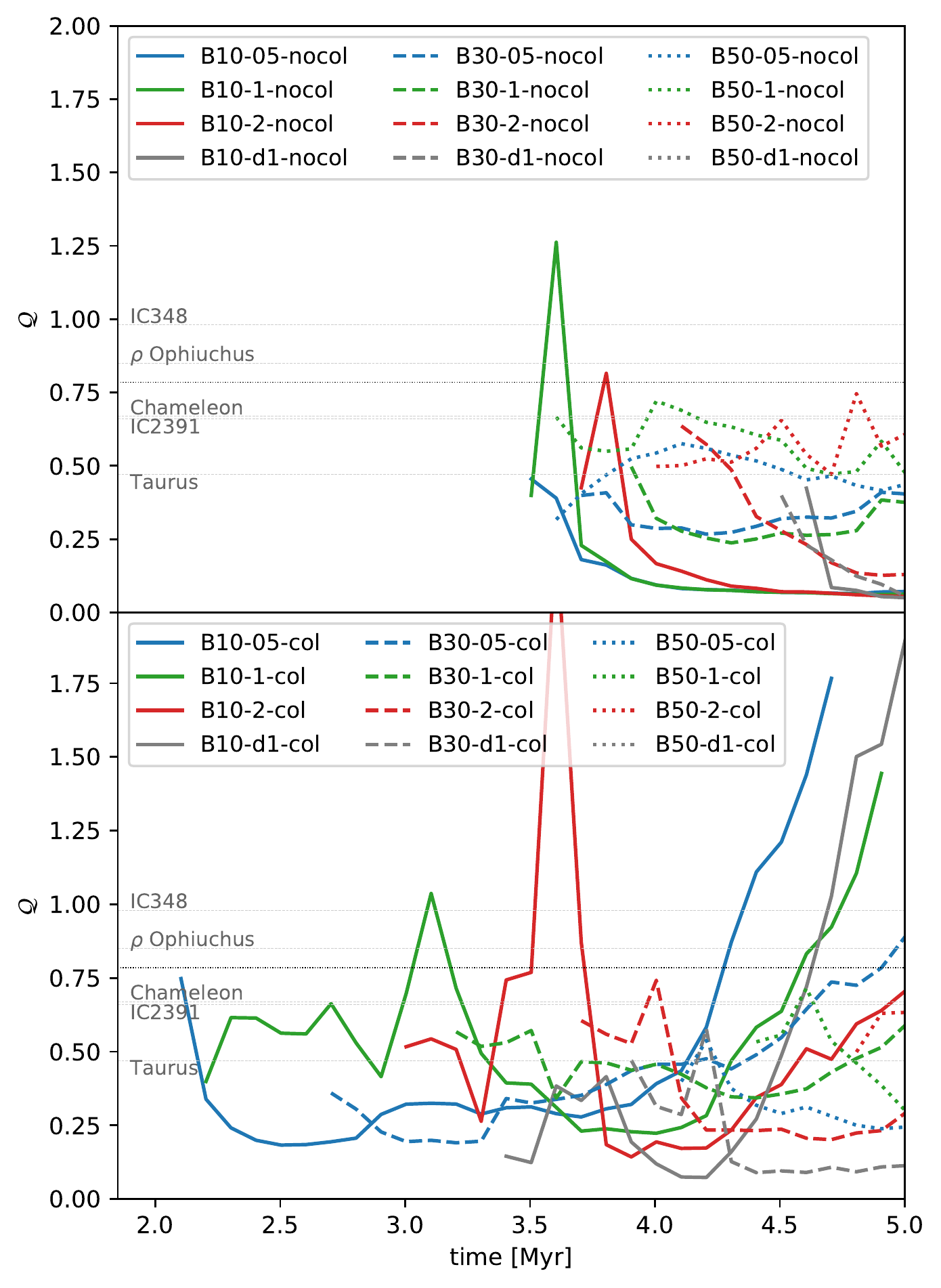}
\caption{
MST $\mathcal{Q}$ parameter versus time. The evolution of $\mathcal{Q}$
is shown for select non-colliding (top) and colliding (bottom)
models. Line colors represent different SF models while line styles
represent different magnetic field strengths.
The values of $\mathcal{Q}$ are averaged over three cardinal
lines of sight in the simulation: $x$, $y$, and $z$. The dotted black
line denotes the threshold of $\mathcal{Q}_{0} = 0.8$ separating
smooth radial versus multi-scale clustering. The gray dashed lines show
values of $\mathcal{Q}$-parameters of various observed star clusters
from \citet{Cartwright_Whitworth_2004}.
\label{fig:Qparam}}
\end{figure}

Figure~\ref{fig:Qparam} shows the evolution of $\mathcal{Q}$ over
time. The non-colliding models show a general trend toward more
multi-scale clustering over time. This is consistent with the
additional star clusters forming throughout different regions in the
GMC complex over time, resulting in a very decentralized overall
distribution of stars. As the strength of the $B$-field is increased,
the cluster distribution becomes less substructured as star formation
occurs in less physically separated regions. Thus, the $\mathcal{Q}$
parameter significantly increases with $B$-field, though it remains
under the $\mathcal{Q}_{0}$ threshold for all non-colliding models.

The colliding models reach higher $\mathcal{Q}$ values than their
non-colliding counterparts, often placing them within the range of
observed star-forming regions. However, after approximately
$t=4.5\:{\rm Myr}$, the dominant cluster in the $B=10\:{\mu G}$
colliding models tends become more and more centrally condensed,
yielding values up to $\mathcal{Q}\sim 2$. This behavior may largely
be due to the lack of stellar feedback, which should limit the
production of additional stars within the cluster potential and likely
produce less gravitationally bound systems that may have smaller
values of $\mathcal{Q}$. The 30 and $50\:{\mu G}$ models show less
fluctuation in the degree of clustering, with higher $B$-fields
generally yielding higher $\mathcal{Q}$.
Sharp maxima in $\mathcal{Q}$ are correlated with increases in the SFR, as new star particles are rapidly created in nearby simulation cells. The inclusion of star particles outside of the main cluster decrease $\mathcal{Q}$.

\subsubsection{Angular Dispersion Parameter}
\label{subsec:ADP}

\begin{figure*}[htb]
\centering
\includegraphics[width=2.1\columnwidth]{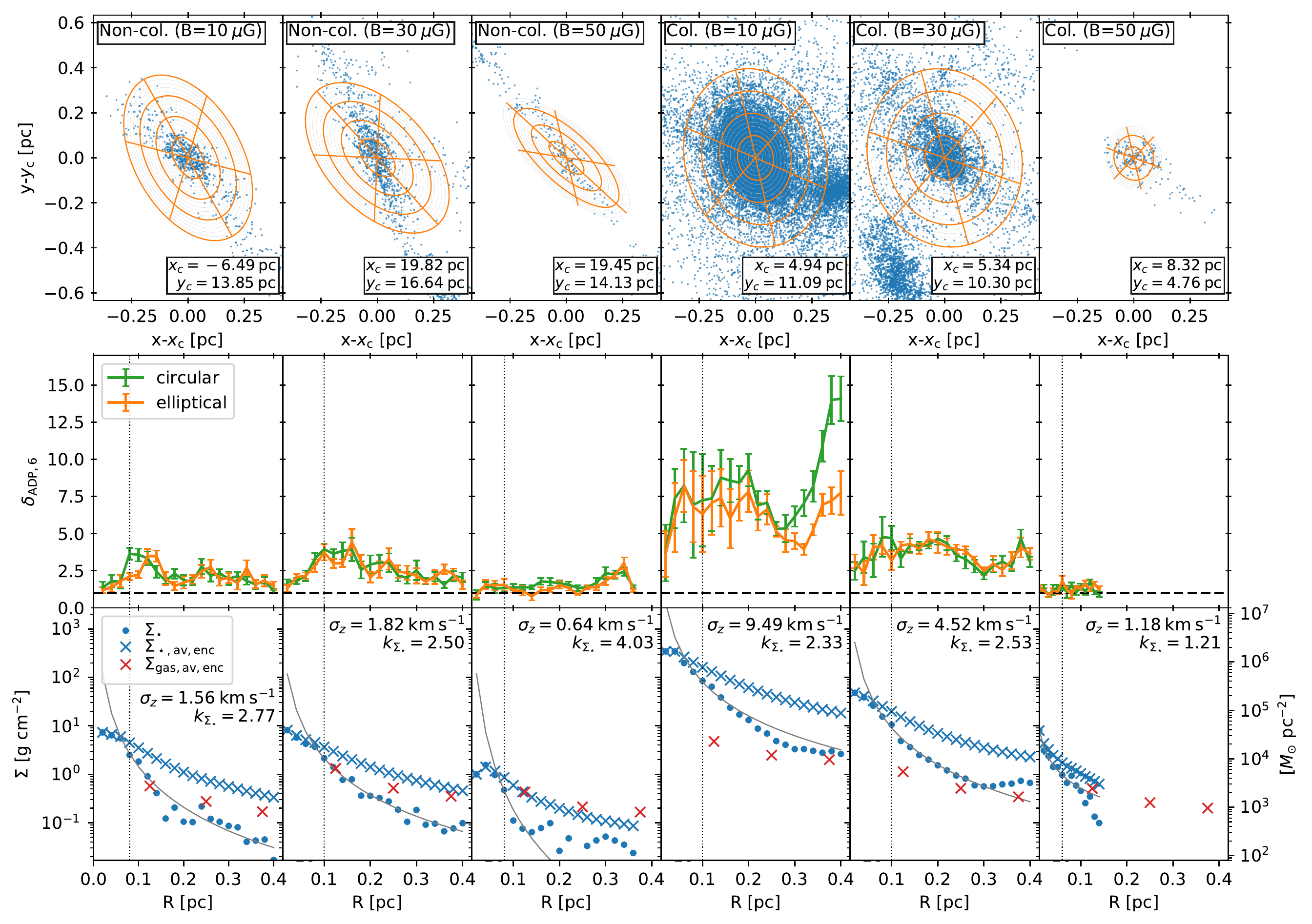}
\caption{
{\it Top row:} The primary cluster in each magnetically-regulated star
formation simulation at $t=5.0$~Myr as defined by DBSCAN. Blue points
denote the projected positions of the star particles, all shown on the
same scale and centered iteratively, with final coordinates
displayed. The elliptical annuli used to calculate the ADP,
$\delta_{\rm ADP,N}$, are drawn in gray, with every fifth annulus and
sector division drawn in orange for clarity. Note that no circular
annuli are drawn, though they are still used in one version of
$\delta_{\rm ADP,6}$ and the radial profiles, both described below.
{\it Middle row:} $\delta_{\rm ADP,6}$ versus radius from the cluster
center. Values are calculated using both the circular (green) and
elliptical (orange) annuli and averaged over twenty $3\degree$
rotations of the 6-sector pattern. The standard error of the mean is
shown by the error bars. The dashed line at $\delta_{\rm ADP,6}=1$
represents a purely random azimuthal distribution of particles (i.e.,
Poisson).  {\it Bottom row:} Stellar and gas mass surface densities as
a function of radius from each primary cluster center. The stellar
profile is shown in blue, with $\Sigma_{\star}$ calculated locally
only within each circular annulus and $\Sigma_{\rm \star, av, enc}$
showing the enclosed average quantity. The gray solid line represents
a power law fit to $\Sigma_{\star}$ with exponent $k_{\Sigma_\star}$
given. The black dotted line denotes the stellar half-mass radius,
$R_{1/2}$. The red crosses indicate the average enclosed gas mass,
$\Sigma_{\rm gas, av, enc}$. The 1-D velocity dispersion of the stars,
$\sigma_{z,1/2}$, is calculated for the cluster as defined by the
half-mass radius.
\label{fig:ADP}}
\end{figure*}

The ADP, $\delta_{\rm ADP, N}(R)$ \citep{DaRio_ea_2014}, is a method
that specializes in quantifying the degree of substructure within a
stellar cluster. It is sensitive to variations in projected cluster
surface density in both the azimuthal and radial directions.  This
method begins by first spatially dividing the distribution of points
among $N$ circular (or elliptical) sectors of equal area. The
dispersion of the number of particles contained within each sector is
calculated, where values near unity for $\delta_{\rm ADP, N}$ indicate
random azimuthal distribution, while higher values of $\delta_{\rm
  ADP, N}$ correspond to more non-uniform, sub-clustered
distributions. Radial dependence can be studied by adding divisions
with concentric annuli. We adopt the best-fitted elliptical annuli to
account for global eccentricity of the cluster using methods identical
to those of Paper III.

The ADP is defined to be
\begin{equation}
    \delta_{\rm ADP, N} = \sqrt{\frac{1}{\left( N-1 \right)\overline{n}} \sum^{N}_{i=1} \left( n_{i}-\overline{n} \right)^{2}} = \sqrt{\frac{\sigma^2}{\sigma^{2}_{\rm Poisson}}}
\end{equation}
for an annulus that is divided into $N$ equal sectors, where $n_{i}$
stars are contained within each $i$th sector. $\overline{n}$ is the
average number of stars per sector within the annulus, $\sigma$ is the
standard deviation of the set of $n_{i}$, and $\sigma_{\rm Poisson}$
is the standard deviation of a Poisson distribution.

In this work, we calculate $\delta_{\rm ADP, N}$ in the primary
cluster for each non-colliding and colliding simulation ($B=10$, 30,
50\:${\rm \mu G}$) using the magnetically-regulated star formation
model with $c_{1}=c_{\rm 1, fid}$. Figure~\ref{fig:ADP} shows results
from both circular and elliptical annuli centered at the iteratively
determined cluster center. We use $N=6$ sectors and typically 20
equally-spaced concentric annuli out to a maximum radius of
0.4\:pc. For the 50\:${\rm \mu G}$ non-colliding and colliding cases,
respectively, we use 18 annuli out to 0.36\:pc and 14 annuli out to
0.14\:pc due to the smaller cluster extents and lower stellar
densities formed in these models. The dispersion is calculated twenty
times using sector patterns rotated every $3\degree$ and averaged to
obtain $\delta_{\rm ADP, 6}$.

Immediately evident is the difference in sizes and distributions of
the most massive cluster formed from quiescent evolution in a
non-collision case compared to those formed in a GMC-GMC
collision. The non-colliding models form relatively elongated primary
clusters along gas filamentary structures, with only a slight
drop-off in membership as the global initial $B$-field increases. On the
other hand, the primary clusters in the colliding models exhibit more
complex substructure in a more crowded environment and experience
large decreases in population as the $B$-field increases.

The ADP as calculated from circular and elliptical annuli do not
differ greatly as functions of radii. $\delta_{\rm ADP, 6}$ for each
of the non-colliding models behaves near Poisson at the center, then
increases to moderate levels of dispersion. The $50\:{\rm \mu G}$ case
shows lesser degrees of dispersion.

In the $10\:{\rm \mu G}$ colliding case, $\delta_{\rm ADP, 6}\simeq
7.5$, indicating a much higher degree of angular dispersion,
especially when a secondary cluster is incorporated near $R=0.4\:{\rm
  pc}$. Clusters in the higher $B$-field cases show lower levels of
dispersion.

The primary clusters formed in the non-colliding models and the more
strongly magnetized colliding models in fact exhibit similar ADP
values as the ONC, where $\delta_{\rm ADP, 6}\simeq 2$
\citep{DaRio_ea_2014}. The $10\:{\rm \mu G}$ colliding case forms a
primary cluster with much higher ADP. Implementing various forms of
local feedback should work to lower the degree of substructure in each
scenario, and the ADP method of quantifying cluster substructure may
be a useful test of such effects.

\subsubsection{Dynamical State of Primary Cluster}
\label{S:clust-props}

The bottom row of Figure~\ref{fig:ADP} shows stellar and gas mass
surface densities as functions of radius. Note that only circular
annuli are used in this analysis. $\Sigma_{\star}$ shows the mass
surface density of stars within each annulus, $\Sigma_{\rm \star, av,
  enc}$ shows the enclosed average quantity at a given radius, while
$\Sigma_{\rm gas, av, enc}$ represents the enclosed average quantity
for gas.

The non-colliding models have relatively similar radial profiles, with
$\Sigma_{\rm gas, av, enc}$ similar to $\Sigma_{\rm \star, av, enc}$
at $\sim1-10\:{\rm g\:cm^{-2}}$ near the cluster center and decreasing
at greater radii. As the global magnetic field is increased, the mass
surface densities decrease, though the fraction of gas to star mass
surface density increases. This can be explained by the corresponding
decrease in level of star formation activity.

Much higher star formation activity is present in the colliding
models, which explains the $\sim1-2$ orders of magnitude higher mass
surface densities. In these cases, $\Sigma_{\rm \star, av, enc}$
exceeds $\Sigma_{\rm gas, av, enc}$ significantly, i.e., achieving
much higher local star formation efficiencies.

A power law is fit to $\Sigma_{\rm \star}$, 
\begin{equation}
    \Sigma_{\rm \star}(R) \propto \left( \frac{R}{0.2\:{\rm pc}} \right) ^{-k_{\sigma_{\star}}},
\end{equation}
where $R$ is the radius from the cluster center and
$k_{\sigma_{\star}}$ is the power law index. This index is generally
about 2.5 in the primary clusters, but varies for more irregular clusters
formed in the $50\:{\rm \mu G}$ models.

The half-mass radius, $R_{1/2}$, is defined as the radius within which
half of the total cluster mass from the maximum aperture radius
(typically 0.4\:pc) is contained. $R_{1/2}$ is generally near 0.1\:pc
for the primary clusters. The stellar masses contained within
$R_{1/2}$ for the non-colliding models are $4.0\times 10^2$,
$5.5\times 10^2$, and $8.4\times 10^1\:{M_{\odot}}$, respectively, for
$B=10$, 30, and $50\:{\rm \mu G}$. For the respective colliding
models, they are $2.2\times 10^4$, $2.7\times 10^3$, and $9.2\times
10^1\:{M_{\odot}}$. The clusters formed in a collision have a much
stronger dependence on the global magnetic environment compared with
those formed in non-colliding models.

Stellar mass surface densities estimated within observed clusters of
similar mass $M_{\star, 1/2}\lesssim 1.0\times10^{3}\:M_{\odot}$ are
generally lower than those found in these simulations
\citep{Tan_ea_2014}, except for the strongest magnetization
cases. Again, this may be explained by the lack of stellar feedback in
our simulations and is expected to better match observations when
protostellar outflow and radiative feedback mechanisms are
implemented.

The dynamical state of a cluster can be estimated using the virial
ratio,
\begin{equation}
    Q = -\frac{T_{\star}}{\Omega} = \frac{3\sigma^{2}R}{2 G M_{\star}},
\end{equation}
where the total kinetic and gravitational potential energies of the
stars are given by $T_{\star}$ and $\Omega$, respectively. $R$ is the
radius that contains a total stellar mass of $M_{\star}$ and $\sigma$
is the 1-D velocity dispersion of all enclosed stars. $Q = 0.5$
indicates a state of virial equilibrium and values above and below 1
indicate clusters that are gravitationally unbound and bound,
respectively. $Q$ can also be related to the virial parameter
\citep{Bertoldi_McKee_1992}, commonly defined as
\begin{equation}
    \alpha_{\rm vir}=3\left( \frac{5-2n}{3-n} \right) \left( \frac{\sigma^{2}R}{G M} \right),
\end{equation}
where $M$ is the total mass enclosed in $R$ and $n$ is the index of the radial density profile, $\rho(r)\propto r^{-n}$. For an $n=2$ profile, this yields a relationship
of $Q=0.5\alpha_{\rm vir}$.

At $R_{1/2}$, virial ratios of 0.17, 0.21, and 0.14 are found for the
non-colliding primary clusters at $B=10$, 30, 50\:${\rm \mu G}$,
respectively. For the colliding primary clusters, respective virial
ratios of 0.14, 0.26, and 0.31 are found. These are all sub-virial,
though the more highly magnetized colliding cases seem to form primary
clusters closer to virial equilibrium.

However, since the half-mass radii can be $\sim 0.1$~pc, which is
about the maximum resolution of AMR grid cells and below the scale at which
gravity is softened, these results are likely to be affected by poor
numerical resolution. Higher resolution studies are needed to
investigate the validity of these results.

\subsection{Gas and Star Kinematics}
\label{sec:results-PV}

\begin{figure*}
  \centering
  \includegraphics[width=1.05\columnwidth]{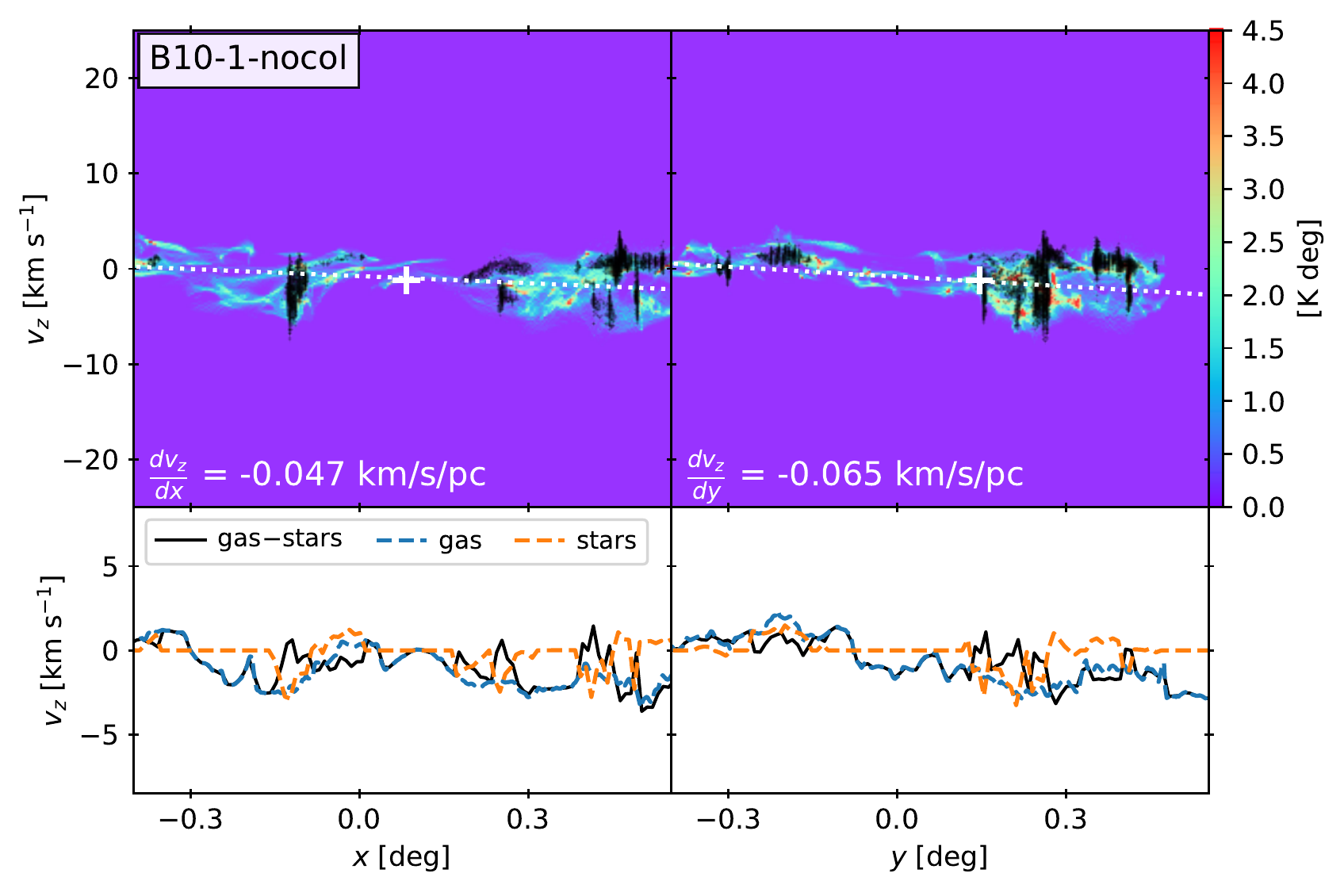}
  \includegraphics[width=1.05\columnwidth]{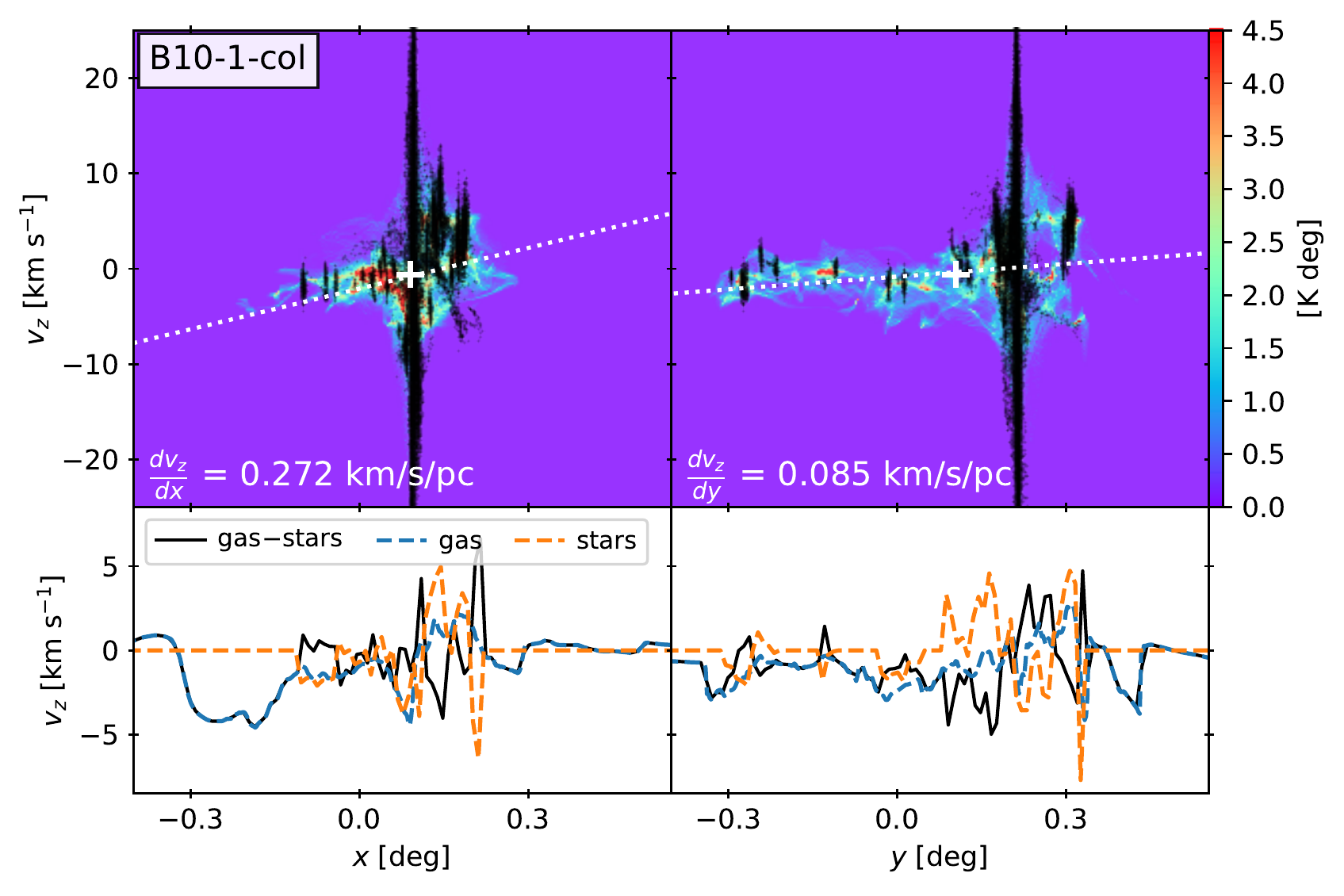}
  \includegraphics[width=1.05\columnwidth]{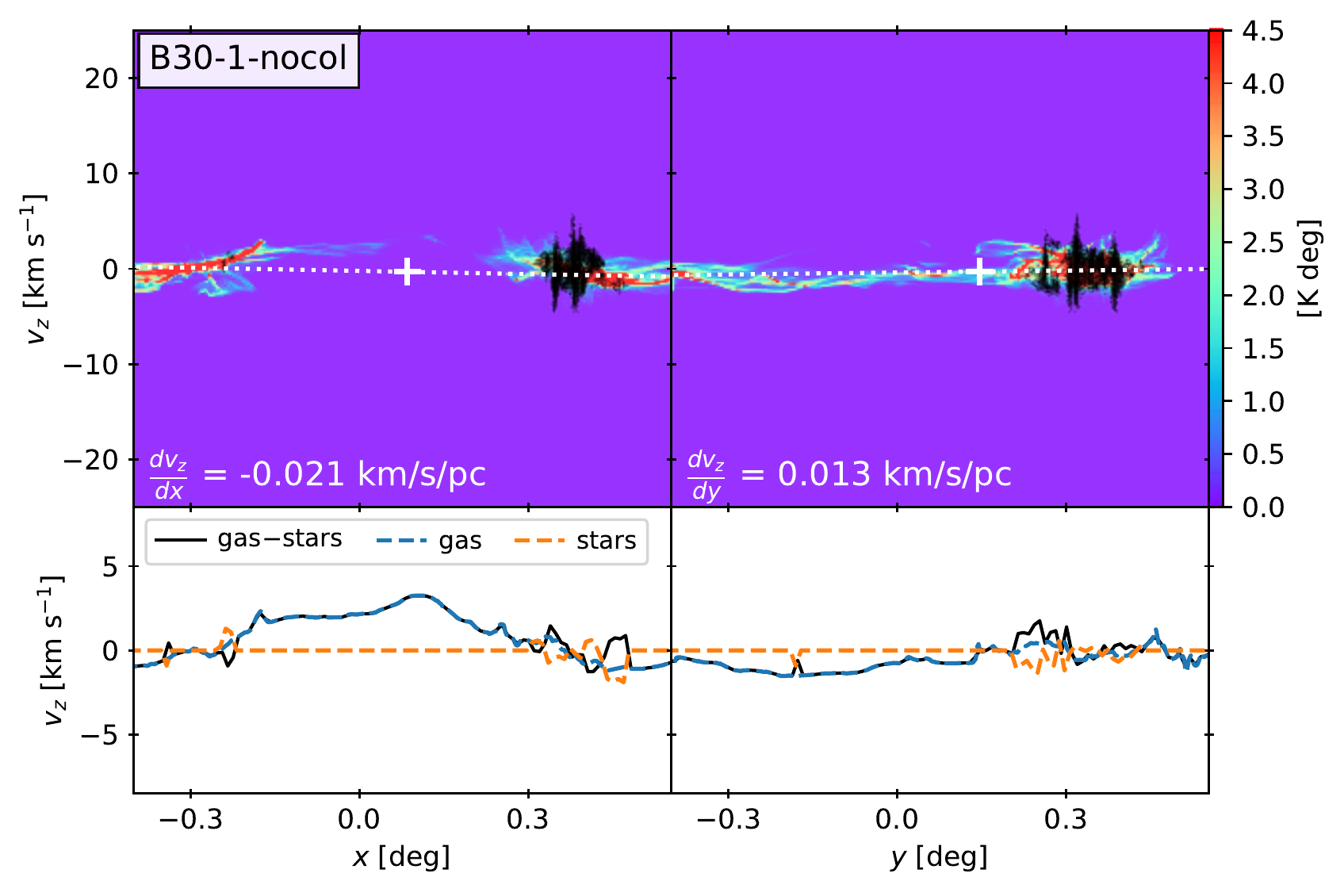}
  \includegraphics[width=1.05\columnwidth]{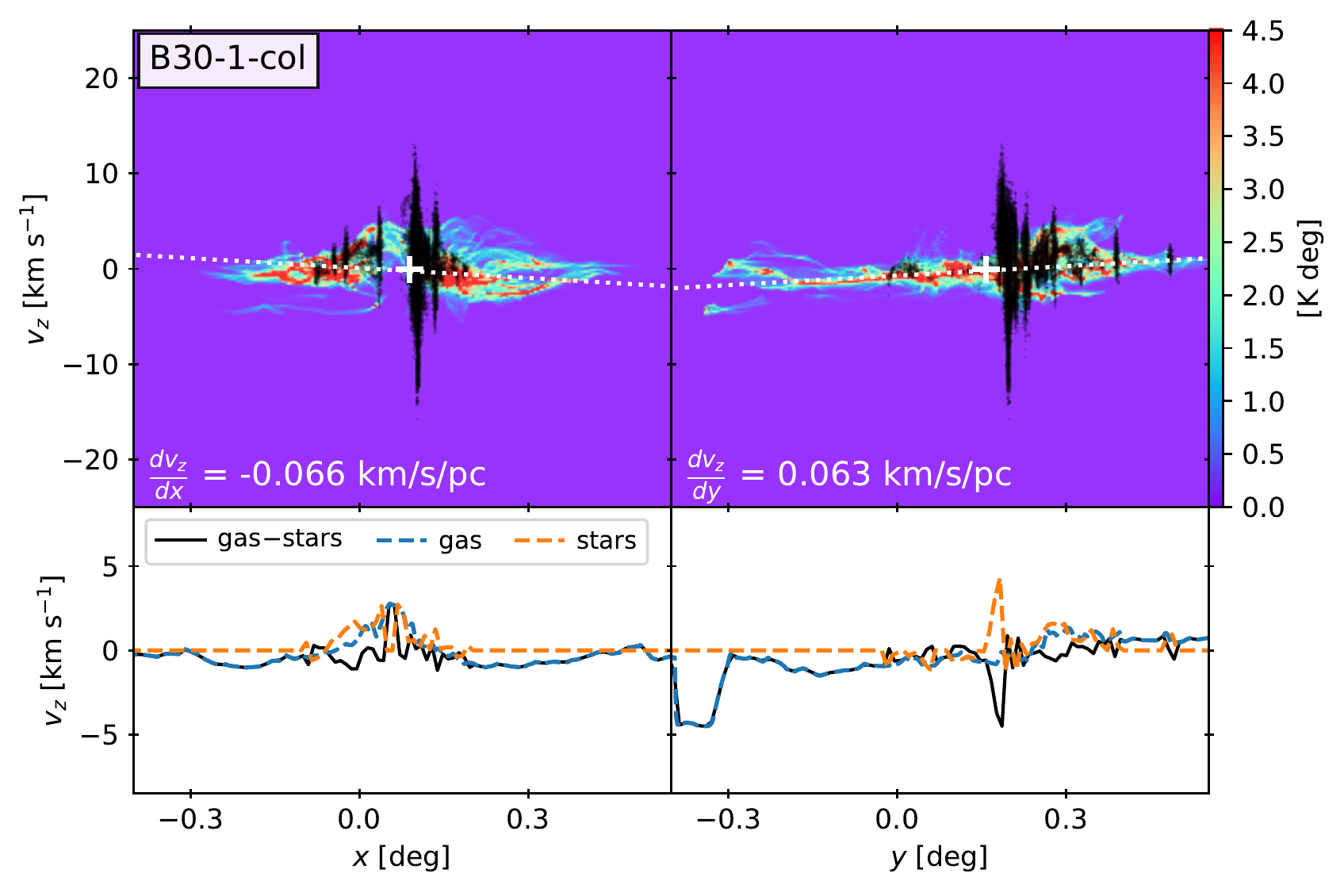}
  \includegraphics[width=1.05\columnwidth]{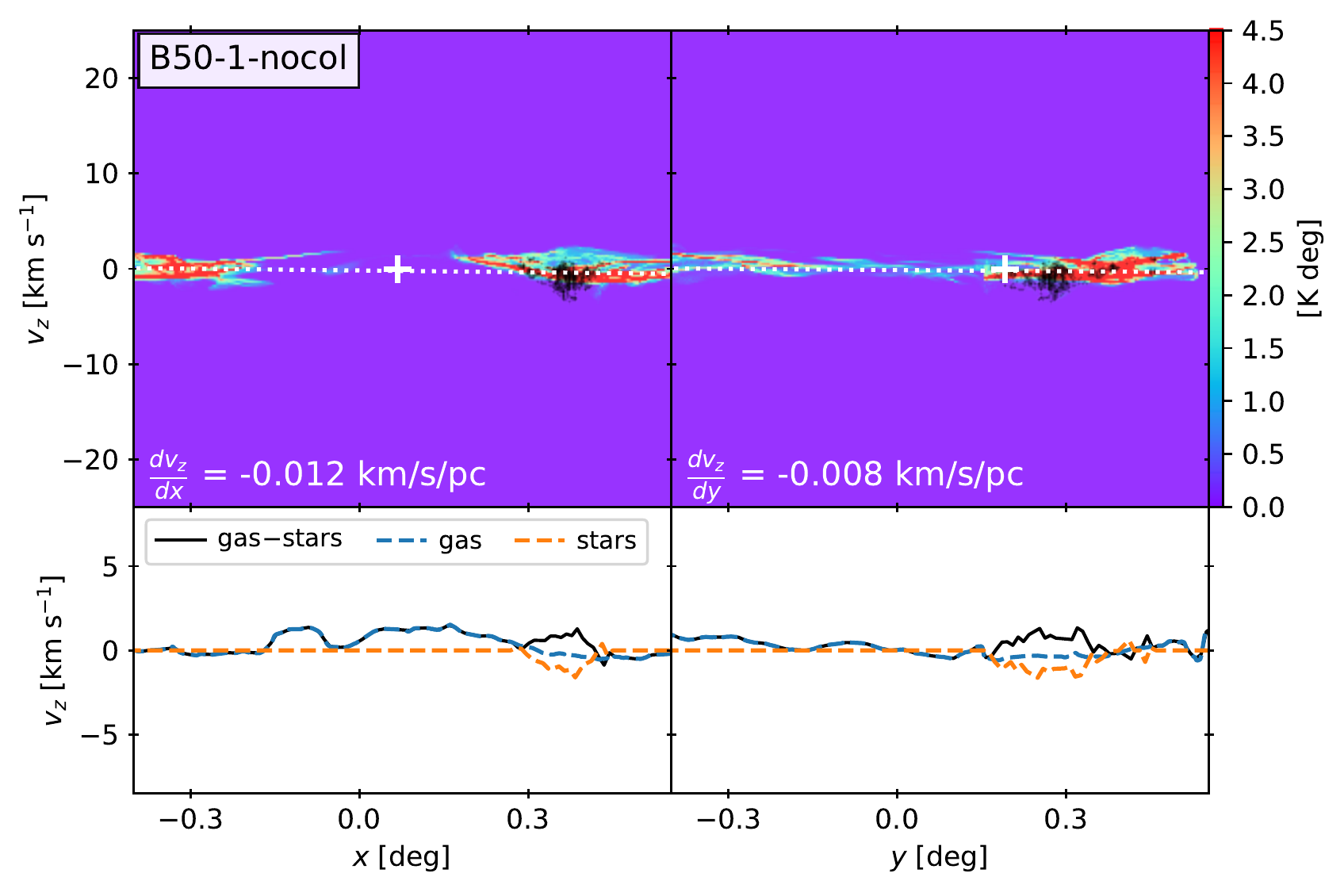}
  \includegraphics[width=1.05\columnwidth]{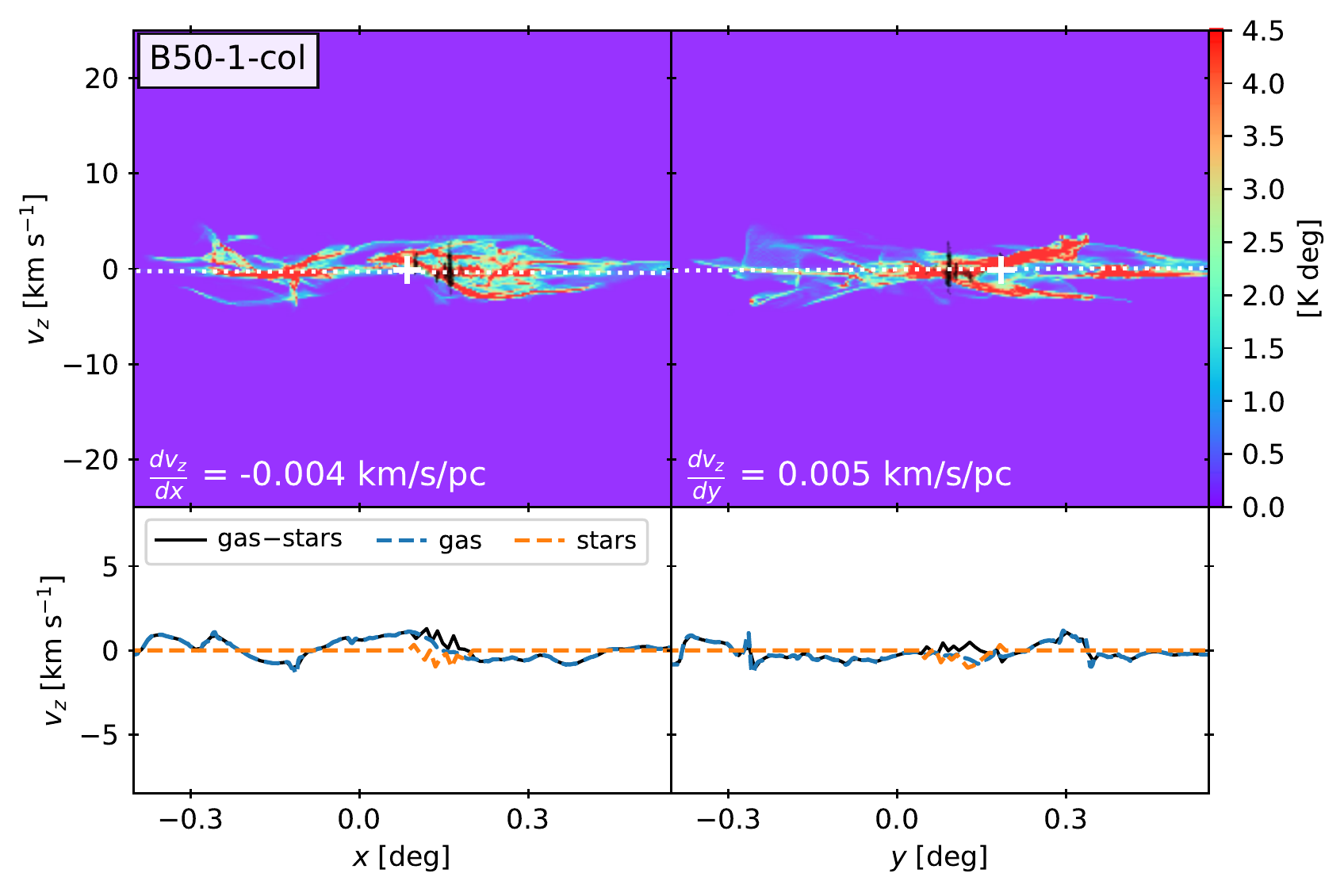}
\caption{
Position-velocity diagrams for non-colliding (left column) and
colliding (right column) simulations for the $B=10\:{\rm \mu G}$ (row
1), $30\:{\rm \mu G}$ (row 2), and $50\:{\rm \mu G}$ (row 3)
simulations using $c_{1}=c_{\rm 1, fid}$. Each model is shown at
$t=5.0\:{\rm Myr}$ along the $z$ line-of-sight.  The colormap depicts
synthetic $^{13}$CO($J$=1-0) line intensities from the gas through
velocity bins of $\Delta v=0.212\:{\rm km\:s^{-1}}$. The star
particles are shown as black points. The gray cross indicates the
position of the center of mass and the solid white line shows the
intensity-weighted linear velocity gradient ($dv_{\rm los}/ds$) across
each cloud. Below each respective position-velocity diagram are plots
of the mean gas velocity, mean star velocity, and their
difference. Positional bins of $0.5~{\rm pc}$ (i.e., $9.5\times
10^{-3}~{\rm deg}$ for an adopted system distance of 3~kpc) are used.
\label{fig:SFppv}}
\end{figure*}

Evidence of the formation mechanism of young embedded clusters may
remain imprinted shortly after on the kinematics of young stars and
their surrounding gas. Position-velocity diagrams have been used to
investigate the kinematics of cluster forming environments in
simulations and observations \citep[see, e.g.,
][]{Duarte-Cabral_ea_2011, Dobbs_ea_2015, Butler_ea_2015,
  Haworth_ea_2015a}. Radial velocity differences, $\Delta v_{r}$,
between young stellar objects and $\rm ^{13}$CO along their line of
sight have been analyzed by \citet{DaRio_ea_2017}, who found that
regions with the greatest differences coincided with the more evolved
regions of the cluster.

Using similar procedures as \citet{Wu_ea_2017b},
Figure~\ref{fig:SFppv} compares position-velocity diagrams between
non-colliding and colliding models for $B=10$, 30 and $50\mu$G,
respectively.  The gas represented by synthetic $^{13}$CO($J$=1-0)
emission has velocity resolution of $0.212~{\rm km\:s^{-1}}$ and is
assumed to be optically thin and located at $d$=3~kpc. Also plotted
are the $^{13}$CO intensity-weighted radial velocities for the gas,
mass-weighted radial velocities of the stars, and the difference
between the two velocity components.

Gas in the non-colliding models is generally dispersed, with moderate
velocity dispersions of a few km/s throughout the various line of
sight positions. As $B$ increases, the velocity dispersion of the gas
and the stars is reduced. The magnitude of the gas velocity gradient
experiences a strong reduction as well. The spatial distribution of
stars changes from numerous small, scattered clusters to one primary
region of star formation. Velocity differences between gas and stars
are generally $\sim 1~{\rm km\:s^{-1}}$, with slight fluctuations.

Collisions trigger the creation of more disrupted gas and star
kinematics that can be readily seen in position-velocity space. The
gas is more concentrated in the collision axis ($x$) due to the
large-scale flows and also in the orthogonal axes ($y$ shown here)
from the higher gravitational potential of the dense clump. The
velocity dispersion of stars and gas is also greatly enhanced,
reaching more than $\pm 5~{\rm km\:s^{-1}}$ in the $B=10\mu$G
case. Cloud collisions also appear to impart higher velocity gradients
in the dense clump. Gas and star velocity differences, too, see large
fluctuations and magnitudes relative to the non-colliding cases. As
$B$ increases, the magnitudes of velocity dispersions, gradients, and
differences are all sharply reduced. However, a higher gas velocity
dispersion relative to the respective non-colliding cases remains
present.

\section{Discussion and Conclusions}
\label{sec:conclusion}

We have presented MHD simulations to investigate how magnetic field
strengths affect the star formation process in self-gravitating,
magnetized, turbulent GMCs evolving relatively quiescently compared
with identical GMCs undergoing collisions at a relative speed of
10~km/s. $B$-field strengths of $B=10$ (i.e., moderately magnetically
supercritical), 30 (i.e., marginally supercritical), and $50\mu$G
(i.e., near critical) are explored, with star formation being
governed by sub-grid routines regulated by either the gas density or
the local mass-to-flux ratio. In such environments, the cloud
and cluster morphology, magnetic field orientations and strengths,
properties of star-forming gas, star formation rates and efficiencies,
and star versus gas kinematics were analyzed.

Stronger $B$-fields are seen to reduce the degree of fragmentation in
both non-colliding and colliding cases and significantly alter the
collision process due to increased magnetic pressure in the
intervening material. The resulting number of stars is reduced and
distributed among fewer clusters.

The relative orientations between filamentary structures and
$B$-fields become increasingly preferentially perpendicular in the
presence of stronger $B$-fields, \edit1{evident morphologically 
as well as quantitatively through HROs}. 
This effect is seen most prominently
in higher column density regions. In low column density regions, there
exists approximately random relative orientations in the non-colliding
models, while the collision forms preferentially parallel
orientations.  Weaker global fields in fact created subregions with
the highest $n_{\rm H}$ and strongest $|B|$, while stronger fields
limited the overall dispersion in density and $B$-field values.

$\Sigma$-PDFs have imprints of the initial $B$-field strength
at early times, where stronger fields form slightly higher-$\Sigma$
distributions in the non-colliding cases, but collisions more greatly
enhance the $\Sigma$ of weaker field cases. At later stages in the
evolution, some $\Sigma$-PDFs develop structures not well-fit by
lognormals.

Stars that formed in the non-colliding models all fell within the $\le
1 M_{\odot}$ regime, and their natal gas exhibited distributions of
density, temperature, magnetic criticality and velocity that narrowed
as $B$-field strength increased. However, the supercritical colliding
models produced a higher dispersion of gas properties, with the
$10\mu$G model in particular forming stars with a distribution of
higher masses in approximate agreement with a power law distribution,
$dN/d\:{\rm log}m_*^{\alpha_*}$ with index $\alpha_*=$-1.35.

For $B=10\mu$G, colliding GMCs resulted in a factor of 10 increase in
SFR and thus a factor of 10 increase in efficiency relative to the
non-colliding counterparts, in agreement with previous
studies. However, this enhancement is reduced and even reversed in
simulations with stronger global fields, which inhibit the collision
and star formation. These results suggest a potential role of cloud
collisions in efficiently forming massive star clusters, while lower
mass star formation may typically take place in more quiescent or more
strongly magnetized environments.

The spatial distribution of star formation, e.g., as measured by the
minimum spanning tree $\mathcal{Q}$ parameter, is much more dispersed
in the non-colliding cases and much more concentrated, i.e., in a more
dominant primary cluster, in the colliding cases. The primary cluster
formed within each simulation showed larger sub-structure in the
colliding cases. All primary clusters were found to be sub-virial, but
the more strongly magnetized colliding cases exhibited clusters
closest to virial equilibrium. However, since the half-mass radii can
be $\sim 0.1$~pc, which is about the maximum resolution of AMR grid
cells and below the scale at which gravity is softened, these results are
likely to be affected by numerical under resolution.

Stronger $B$-fields also resulted in reduced velocity dispersions,
velocity gradients, and degree of stellar kinematics overall. Such
properties were enhanced by the collision in each $B$-field case.

\acknowledgments B.W. would like to acknowledge financial support by a Grant-in-Aid
for Scientific Research (KAKENHI Number 18H01259) of Japan Society for the Promotion of Science (JSPS). Computations described in this work were performed using the publicly-available \texttt{Enzo} code (http://enzo-project.org). This research also made use of the yt-project (http://yt-project.org), a toolkit for analyzing and visualizing quantitative data \citep{Turk_ea_2011}.  
The authors acknowledge University of Florida Research Computing (www.rc.ufl.edu) and the Center for Computational Astrophysics at NAOJ for providing computational resources and support that have contributed to the research results reported in this publication. 

\software{Enzo \citep{Bryan_ea_2014}, Grackle \citep{Kim_ea_2014}, PyPDR, yt \citep{Turk_ea_2011}}


\clearpage

\end{document}